\documentclass{aastex61}

\usepackage{amssymb}

\def\eg{{\it e.g.,} }
\def\etal{{\it et al. }}
\def\ie{{\it i.e.,} }

\def\rs{r_{\rm S}}
\def\mbh{M_{\rm {bh}}}
\def\vr{v_{\rm r}}
\def\vre{v_{\rm re}}
\def\vrr{v_{\rm r2}}
\def\vrt{v_{\rm r1}}
\def\vt{v_{\rm{\theta}}}

\def\vtr{v_{\rm \theta 2}}
\def\vtt{v_{\rm \theta 1}}
\def\vp{v_{\rm \phi}}
\def\vpe{v_{\rm \phi e}}
\def\vpr{v_{\rm \phi 2}}
\def\vpt{v_{\rm \phi 1}}
\def\Thr{\Theta_2}
\def\Thre{\Theta_{\rm e}}
\def\Thth{\Theta_1}
\def\rh{\rho_2}
\def\rhe{\rho_{\rm e}}
\def\rht{\rho_1}
\def\trp{\tau_{\rm r\phi}}
\def\trpe{\tau_{\rm r\phi_e}}
\def\ttp{\tau_{\rm\theta\phi}}
\def\omegk{\Omega_{\rm K}}

\def\lsim{\lower.5ex\hbox{$\; \buildrel < \over \sim \;$}}
\def\gsim{\lower.5ex\hbox{$\; \buildrel > \over \sim \;$}}
\def\simeq{\lower.3ex\hbox{$\; \buildrel \sim \over - \;$}}
\def\rin{\rm {in}}

\def\Pe{P_{\rm e}}
\def\epse{\epsilon_{\rm e}}
\def\sun{M_\odot}

\def\rin{r_{\rm in}}
\def\tt{\tilde{t}}
\def\geff{\gamma_{\rm eff}}
\def\lmda{\lambda_{\rm e}}
\def\be{B_{\rm e}}
\def\ase{a_{\rm se}}
\submitjournal{ApJ}

\shorttitle{Inflow-outflow around black holes}
\shortauthors{Kumar \& Gu}

\begin{document}

\title{The 2D disk structure with advective transonic inflow-outflow solutions around black holes}
\email{kumar@xmu.edu.cn, guwm@xmu.edu.cn}

\author{Rajiv Kumar}
\affil{Department of Astronomy, Xiamen University, Xiamen,
Fujian 361005, China}

\author{Wei-Min Gu}
\affiliation{Department of Astronomy, Xiamen University, Xiamen,
Fujian 361005, China}
\affil{Jiujiang Research Institute of Xiamen University, Jiujiang 332000, China}



\begin{abstract}
We solved analytically viscous two-dimensional (2D) fluid equations for accretion and outflows in 
spherical polar coordinates ($r, \theta, \phi$) and obtained explicitly 
flow variables in $r-$ and $\theta -$directions around black holes (BHs). 
We investigated global transonic advection-dominated accretion flow (ADAF) solutions in $r-$direction  
on an equatorial plane with using Paczy\'nski-Wiita potential. 
We used radial flow variables of ADAFs with symmetric conditions on the equatorial plane, as initial 
values for integration in $\theta-$direction. 
In the study of 2D disk structure, we used two-azimuthal components of viscous stress tensors namely, $\trp$ and $\ttp$.
Interestingly, we found that the whole advective disk is not participating in outflow generation and
the outflows form close to the BHs.  
Normally, outflow strength increased with increasing viscosity parameter ($\alpha_1$), mass-loss parameter ($s$) 
and decreasing gas pressure ratio ($\beta$).  
Outflow region increased with increasing $s$, $\alpha_1$ for $\trp$ and decreasing $\alpha_2$ for $\ttp$.  
The $\ttp$ is effective in angular momentum transportation at high latitude and 
outflows collimation along an axis of symmetry 
since it changes polar velocity ($\vt$) of the flow.
The outflow emission is also affected by the ADAF size and decreased with decreasing it. 
Transonic surfaces formed for both inflows ($\vr<0$, very close to BH) and outflows ($\vr>0$). 
We also explored no outflows, outflows and failed outflows regions, 
which mainly depend on the viscosity parameters.
\end{abstract}
\section{Introduction} \label{sec:intro}
An accretion disk is associated with many astrophysical objects, \eg compact objects (black holes, neutron stars,  
and white dwarfs) and young stellar objects. 
Accreting gas onto these objects can generate radiation and bipolar outflows/jets due to an extraction of its gravitational energy. 
These objects are often associated with non-relativistic to relativistic bipolar jets. Especially, the relativistic
jets have been observed around accreting black hole candidates (BHCs), for instance, active galactic
nuclei (AGNs) and black hole $X-$ ray binaries (BHXBs). The jets from the AGN M87 are emerged from an extremely 
small central region of a source within $100\rs$ \citep{jbl99} but 
recent observation shows that even smaller region less than $10\rs$ \citep{detal12}, 
where $\rs=2G\mbh/c^2$ is a Schwarzschild radius.
AGNs and BHXBs are believed to harbor supermassive BHs $\sim 10^{6-9}\sun$ and stellar mass BHs $\sim 10\sun$ at the center, 
respectively. $\sun$ denotes mass of the Sun. 
Moreover, the BHXBs are also showing typically two type of spectral states in their observations, one high soft state, 
which is radiatively efficient and dominated by thermal radiation with black body spectrum in soft X-ray regime and 
second low hard state, which is radiatively inefficient and dominated by non-thermal radiation with some power 
law spectrum in the hard X-ray regime \citep{rm06}. 
These two states are also connected with many intermediate states and interestingly, bipolar jets and 
quasi-periodic oscillations are associated 
with the hard state in the BHXBs \citep{gfp03,fbg04}. 
However, such changes in the spectral states are yet to be observed for the AGNs. 
Since the time scales of AGNs and BHXBs can be scaled by the mass of the BHs, 
but inner boundary conditions are same, 
so the basic physics of both kind of objects can be similar \citep{mkkf06}.   
In this context, there are a several theoretical and numerical studies on accretion processes with Keplerian/sub-Keplerian flows 
and that can play an important role in generations of soft spectrum \citep{ss73,nt73,acls88} and 
hard spectral state \citep{st80,ct95,ny95b,emn97}, hard/soft state transitions \citep{wxyg16}, 
and the outflows from the accretion disks around BHs 
\citep{ny95a,mrc96a,msc96b,ia99,ia00,omnm05,otm07,yyob14,dcnm14,bygy16a,bygy16b,lckhr16,kc17,jsd17}. 
The mechanism for the jet generation and evolution of it is still not much clear and 
a topic of active research in the fields of theory and observations.

The analytical study of the 2D disk with outflows has been started with a relaxation of vertical hydrostatic equilibrium 
in the disk by \cite{ny95a}. They used self-similar ADAF solutions \citep{ny94} in the radial direction with 
symmetry conditions on the equatorial plane and solved the flow variables along the polar direction. 
However, they could not get actual outflow solutions because they assumed mass accretion rate is independent of 
radial distance and thus polar velocity, $\vt=0$. Since they have found in their solutions that the Bernoulli parameter 
is positive, therefore the outflows may form close to a rotation axis. 
Subsequently, \cite{xc97} have included the $\vt$ non-zero in their study and found accretion and ejection solutions. 
After that, the theoretical studies of the 2D disk with self-similar solutions along the radial directions have been done 
by many authors with the outflows \citep{XW05,gxll09,JW11,g12,g15} and without outflows \citep{has17}. 
The similar studies have also done in magneto-hydrodynamics regime \citep{sa16,mby16,zmay18}.
An one more analytical study for the outflows with self-similar solutions in one-dimensional flow has done by \cite{BB99}
with assuming the mass accretion rate varies with some power of the radial distance. These outflow solutions are known as adiabatic inflow-outflow solutions (ADIOS). 
Further, they have presented their work with a family of two-dimensional self-similar solutions for the outflows \citep{BB04}.
Simultaneously, many numerical simulations for the investigations of accretion-ejection have been also done with optically thin hot flows by 
\cite{spb99,ia99,ia00,ywb12,ybw12,bygy16a,bygy16b}, a review by \cite{yn14} and with optically thick, hyper accreting flows by \cite{omnm05,otm07,yyob14,jsd14,Jetl15,jsd17}.
There are a few more models for explanation of jet generation by the extraction of rotational energy of Kerr BHs \citep{bz77}, 
by anchoring of matter with magnetic lines \citep{bp82} and by shocked generated extra-thermal gradient force in 
the post-shock region \citep{mrc96a,msc96b,kc13,ck16,lckhr16,kc17}.
 
The self-similar solution gets popularity for the analytical studies of the 2D disk with/without outflows in both hydrodynamics (HD) 
and magneto-hydrodynamics (MHD) regime   
because it simplifies fluid equations in the radial direction, which makes ODEs with only polar derivatives or independent of radius and radial derivatives. 
\cite{g12,g15} have mentioned that the outflows form naturally from advective accretion disk for both optically thin and thick gas medium.
Moreover, the hot flows with bremsstrahlung, synchrotron emissivity and Comptonization of soft photons can give rise to the hard spectrum of the BHCs \citep{ct95,ny95b,yn14}.
Since the bipolar jets have been seen during hard state with radiatively inefficient flows around the BHCs \citep{rm06}. 
Therefore, we assumed inner part of the disk is hot advective and radiatively inefficient accretion flows for the study of jet generations around the BHs. 
We believed on the two zone configuration of the accretion disk \citep{emn97,ds13}, 
one inner part hot sub-Keplerian advective radiative inefficient accretion flows (RIAFs) and 
other outer part, which is geometrical thin and cool Keplerian optically thick (Shakura-Sunyaev disk).  
For time being, we did not consider radiative emissivities in the flow since this work mainly focus on 
the study of jet generation in the 2D disk and full consideration of radiative advective flows leave for future work.  

The present paper is based on the study of structure of accretion disk in the 2D flow with outflows and 
extension of previous studies \citep{ny95a,xc97,BB99,XW05,xy08,JW11}.
In this paper, there are following things, which are new from the previous 2D analytical studies as one by one,
since we want to study inflow-outflow structure close to the BH, therefore we used pseudo-Newtonian potential \citep{pw80}. 
Which incorporates general relativistic effects very close to the BH. 
Second, we considered two-azimuthal components of viscous stress tensor \citep{spb99,ywb12} from out off nine-components \citep{XW05}. 
{Since it is mostly believed that the azimuthal component of magnetic stress is more important in the 
angular momentum transfer 
by the study of magneto-rotational instability (MRI) simulations \citep{bh98}.
The present study is axisymmetric with the 2D HD rotating flow so 
we used anomalous shear stress, which can approximates the magnetic stress and 
following \cite{spb99,ywb12} for assuming two-azimuthal components of the viscosity are non-zero. 
Their effects on the disk structures have also discussed in the simulation by \cite{yyob14}.}
Third, we have investigated global transonic ADAF solutions on the equatorial plane and 
their flow variables used as the boundary conditions for the integration of differential equations in $\theta-$direction. 
When doing this, the fluid ODEs are still dependent on the radius of the disk and other radial derivatives, 
unlike with using the self-similar assumptions \citep{XW05,JW11}.
In present study, our main interest is to investigate the inflow-outflow structure close to the BHs with changing various flow
parameters, namely, disk viscosity parameters ($\alpha_1~\& ~\alpha_2$), grand specific energy (${E}$),  
{gas pressure ratio ($\beta$)} 
and mass-loss parameter ($s$) in the fluid flows.
The structure of this paper is in next section 2, model fluid equations and assumptions, section 3, solution procedure,
section 4, numerical results and in the last section, conclusions of our study.

\section{Model fluid Equations and Assumptions} \label{sec:eqns}
We considered viscous hydrodynamic fluid equations for advective accretion-outflow solutions with steady-state and 
axisymmetric in the spherical 
polar coordinates ($r, \theta, \phi$). We assumed pseudo-Newtonian geometry 
$\Phi=-G\mbh/(r-\rs)$\citep{pw80} around the Schwarzschild BHs.
For time being, we are ignoring magnetic field in the accretion and outflows.
We represented the viscous fluid equations and the flow variables in geometrical unit
system and chosen $2G=\mbh=c=1$, where, $\mbh, G$ and $c$ are mass of the BH, universal
gravitational constant and speed of the light, respectively.
Therefore, units of a length, flow velocity ( or sound speed), energy, specific angular momentum, mass, density, pressure
and time are $G\mbh/c^2$, $c$, $\mbh c^2$, $G\mbh/c$,  $\mbh$, $c^6/(8G^3\mbh^2)$, $c^8/(8G^3\mbh^2)$ and $G\mbh/c^3$, respectively. 
We also assumed that the two-components of viscous stress tensor are effective in $r-\theta$ plane, which are $\trp$ and $\ttp$ as following \cite{spb99}.
Thus, the conserved form of the fluid equations in the 2D become as,\\ 
the continuity equation,
\begin{equation}
 \frac{1}{r^2}\frac{\partial}{\partial r}(r^2\rho \vr)+\frac{1}{r {\rm sin\theta}}\frac{\partial}{\partial\theta}(\rho \vt {\rm sin\theta})=0,
 \label{ce.eq}
\end{equation}
the components of Navier-Stokes equation,
\begin{equation}
 \vr \frac{\partial \vr}{\partial r}+\frac{\vt}{r}\frac{\partial \vr}{\partial\theta}-\frac{\vt^2+\vp^2}{r}+
 \frac{1}{\rho}\frac{\partial P}{\partial r}-F_{\rm r}=0,
 \label{rc.eq}
\end{equation}
\begin{equation}
 \vr \frac{\partial \vt}{\partial r}+\frac{\vt}{r}\frac{\partial \vt}{\partial\theta}+\frac{\vr \vt}{r}-
 \frac{\vp^2{\rm cot\theta}}{r}+\frac{1}{r\rho}\frac{\partial P}{\partial\theta}=0,
 \label{tc.eq}
\end{equation}
\begin{equation}
 \vr \frac{\partial \vp}{\partial r}+\frac{\vt}{r}\frac{\partial \vp}{\partial\theta}+\frac{\vp}{r}(\vr+\vt {\rm cot\theta})=
 \frac{1}{\rho r}\left[\frac{1}{r^2}\frac{\partial}{\partial r}(r^3t_{\rm r\phi})+\frac{\partial t_{\rm\theta\phi}}{\partial \theta}+
 2t_{\rm\theta\phi}{\rm cot\theta}\right],
 \label{pc.eq}
\end{equation}
and the energy equation,
\begin{equation}
 \rho\left[\vr\frac{\partial\epsilon}{\partial r}+\frac{\vt}{r}\frac{\partial\epsilon}{\partial\theta}-
 \frac{P}{\rho}\{\frac{\vr}{\rho}\frac{\partial\rho}{\partial r}+
 \frac{\vt}{r\rho}\frac{\partial\rho}{\partial\theta} \}\right]=fQ^+,
 \label{eg.eq}
\end{equation}
where $Q^+=\trp^2/{\eta_1}+\ttp^2/{\eta_2}$ is viscous heating rate and $f$ is advection factor \citep{ny95a}. 
For simplicity we assumed $f$ is fixed. Since the values of $f$ should not be arbitrary, therefore for brevity 
we used only $f=1$ for highly advective flow, inspite of radiation-dominated or gas-dominated flow. 
However, $f$ should be determined with relevant radiation mechanisms. 
$P$($=p_{\rm g}+p_{\rm rad}$) is total pressure, $p_{\rm g}={\rho\Theta}/{\tilde{t}}$ is gas pressure and 
$p_{\rm rad}$ is radiation pressure, which could be due to blackbody emissivity for optically thick medium \citep{acls88} or 
bremsstrahlung and synchrotron emissivity for optically thin medium \citep{ny95b}. 
$\Theta=k_{\rm B}T/(m_{\rm e}c^2)$ is dimensionless temperature of the fluid, $\tilde{t}=\mu m_{\rm p}/m_{\rm e}$, 
where $k_{\rm B}$, $\mu$, $m_{\rm p}$ and $m_{\rm e}$ are the Boltzmann constant, mean molecular weight of the gas, 
mass of the proton and mass of the electron, respectively. 
We assumed $\mu =0.5$ for fully ionized flow.
$F_{\rm r}=-d\Phi/dr$ is central attractive force around the BH.
The specific internal energy \citep{kfm08,JW11} is
\begin{equation}
 \epsilon=\frac{p_{\rm g}}{\rho(\gamma-1)}+3\frac{p_{\rm rad}}{\rho}=\frac{P}{\rho(\geff-1)}~~\mbox{(or EoS)},
 \label{eos.eq}
\end{equation}
where $\gamma$ is known as adiabatic index and defined as ratio between heat capacities. 
$\geff=1/[N\beta+3(1-\beta)]+1$ is effective $\gamma$, $N=1/(\gamma-1)$ is polytropic index and $\beta=p_{\rm g}/P$ is 
the gas pressure ratio. 
The two-azimuthal components of the viscous stress tensors 
are written as,
\begin{equation}
 \trp=\eta_1\left(\frac{\partial \vp}{\partial r}-\frac{\vp}{r}\right)~~\mbox{and}~~
 \ttp=\frac{\eta_2}{r}\left(\frac{\partial \vp}{\partial\theta}-\vp {\rm cot\theta}\right),
 \label{vst.eq}
\end{equation}
where $\eta_1={\alpha_1 P}/{\omegk}$ and $\eta_2={\alpha_2 P}/{\omegk}$ are coefficients of viscosity 
and $\omegk={1}/{(\sqrt{2r}(r-1))}$ is Keplerian angular velocity on the equatorial plane. 
The $\alpha_1$ and $\alpha_2$ are the Shakura-Sunyaev viscosity parameters. 
The flow variables in the $r-\theta$ plane are defined \citep{XW05} as:
$$
\mbox{Mass density}~~\rho(r,\theta)=\rho=\rht(\theta)\rh(\theta=\pi/2,r),
$$
$$
\mbox{Radial velocity}~~\vr(r,\theta)=\vr=\vrt(\theta)\vrr(\theta=\pi/2,r),
$$
$$
\mbox{Polar velocity \deleted{or evaporation velocity}}~~\vt(r,\theta)=\vt=\vtt(\theta)\vtr(\theta=\pi/2,r),
$$
$$
\mbox{Azimuthal velocity}~~\vp(r,\theta)=\vp=\vpt(\theta)\vpr(\theta=\pi/2,r),
$$
\begin{equation}
\mbox{Fluid temperature}~~\Theta(r,\theta)=\Theta=\Thth(\theta)\Thr(\theta=\pi/2,r), 
 \label{flwvrt.eq}
\end{equation}
where the flow variables with `$\theta$' in brackets are represented variation along the $\theta-$ direction for a given $r$ and 
they are called as polar flow variables and corresponding derivatives will be polar flow derivatives.
The flow variables with `$r$' in brackets are represented variation along the radial direction 
and they are called as radial flow variables and corresponding derivatives will be radial flow derivatives. 
{Here $\vtr=1/\sqrt{2r}$, we are following same as in the previous studies \citep{XW05,JW11} and 
corresponding radial derivative.} 
Using above definitions in equations (\ref{ce.eq}-\ref{eg.eq}) then we get ordinary differential equations (ODEs) 
of the 2D flows, 
\begin{equation}
 \rho \vr r\left[\frac{2}{r}+\frac{1}{\rh}\frac{d\rh}{d r}+\frac{1}{\vrr}\frac{d \vrr}{d r}\right]+
\rho\vt\left[\frac{1}{\rht}\frac{d\rht}{d\theta}+\frac{1}{\vtt}\frac{d \vtt}{d\theta}+{\rm cot\theta} \right]=0
\label{ce2.eq}
\end{equation}
\begin{equation}
 \vrt\vr\frac{d\vrr}{d r}+\vt\frac{\vrr}{r}\frac{d\vrt}{d\theta}-\frac{v_\theta^2+v_\phi^2}{r}+\frac{\Theta}{\tilde{t}\beta\rh}\frac{d\rh}{dr}
 +\frac{\Thth}{\tilde{t}\beta}\frac{d\Thr}{dr}-F_r=0,
 \label{rc2.eq}
\end{equation}
\begin{equation}
 \vr r\vtt\frac{d \vtr}{dr}+\vt\vtr\frac{d\vtt}{d\theta}+
v_rv_\theta-v_\phi^2{\rm cot\theta}+\frac{\Theta}{\beta\tt}\frac{1}{\rht}\frac{d\rht}{d\theta}+
\frac{\Thr}{\beta\tt}\frac{d\Thth}{d\theta}=0
\label{tc2.eq}
\end{equation}
\begin{eqnarray}\nonumber
 r\vpt\vr\frac{d\vpr}{dr}+\vp(\vr+\vt {\rm cot\theta})+\vt\vpr\frac{d\vpt}{d\theta}=\\
 \frac{\Thth\vpt}{\rh r^2}\frac{d(r^2\trpe)}{dr}+\frac{\alpha_2\vpr\Theta}{\beta\tt\omegk r}\left[\frac{\tau_{\theta}}{\Thth}\frac{d\Thth}{d\theta}+
 \frac{\tau_{\theta}}{\rht}\frac{d\rht}{d\theta}+\frac{d^2\vpt}{d\theta^2}+\vpt+\tau_{\theta}cot\theta\right]
\label{pc2.eq}
\end{eqnarray}
\begin{equation}
 \vr\Theta\left[\frac{N_{\rm eff}}{\Thr}\frac{d\Thr}{dr}-\frac{1}{\rh}\frac{d\rh}{dr}\right]+\frac{\vt\Theta}{r}\left[\frac{N_{\rm eff}}{\Thth}\frac{d\Thth}{d\theta}-
 \frac{1}{\rht}\frac{d\rht}{d\theta}\right]=\beta\tt fQ^+,
 \label{eg2.eq}
\end{equation}
where $\tau_{\rm \theta}=(d\vpt/d\theta-\vpt {\rm cot\theta})$ and $N_{\rm eff}=1/(\geff-1)$ is effective polytropic index.
We have solved above equations (\ref{ce2.eq}-\ref{eg2.eq}) explicit way and following similar methodology as used in papers \citep{XW05,JW11}. Since we are 
avoiding self-similar solution definitions along the radial direction, therefore firstly, 
we have to find out the radial flow variables with corresponding derivatives of the ADAF on the equatorial plane 
(detail equations are presented in appendix \ref{sec:Feqneq}), then get polar flow variables using symmetric boundary conditions on the equatorial plane 
and finally integrate above equations along the polar direction. 
Before doing so, we are assuming some symmetric properties with boundary conditions in the next subsection.
\subsection{Boundary conditions for inflow-outflow}\label{subsec:bcio}
In order to solve ODEs (\ref{ce2.eq}-\ref{eg2.eq}) in $\theta-$direction, so we used symmetric boundary conditions at $\theta=\pi/2$ from the rotation axis. 
Which are obtained from the reflection symmetry and
following the previous studies \citep{XW05,JW11},
\begin{equation}
 \vtt(\pi/2)=0=\frac{d\rht(\pi/2)}{d\theta}=\frac{d\Thth(\pi/2)}{d\theta}=
 \frac{d\vrt(\pi/2)}{d\theta}=\frac{d\vpt(\pi/2)}{d\theta};~~\rht(\pi/2)=1.
 \label{bc.eq}
\end{equation}
{Since $\vt$ is an evaporation velocity for the generation of outflows so we assumed before the outflow at $\theta=\pi/2$, 
it is zero but becomes non-zero immediately,   
when matter goes upward from the equatorial plane. Therefore we used $d\vtt(\pi/2)/d\theta$ is non-zero on the equatorial plane 
and represented below in equation (18). Moreover, the outflows are started from the equatorial plane so here we assumed total flow 
density at $\pi/2$ is equal to the inflow density means $\rho=\rh$, which implies $\rht(\pi/2)=1$ at all the radius. 
Here the $\rh=\rhe$ is changing with the radius and also depends on $\theta_e$ and mass accretion rate ($\dot{m}$) 
as represented in below equation (\ref{mloss.eq}).
But the disk structure is independent of $\theta_e$ and $\dot{m}$.}
By using above definitions, we obtained explicitly the fluid equations in the pure radial direction 
at $\theta=\pi/2$ (appendix \ref{sec:Feqneq}). 
Since gas can evaporate from the accretion disk to infinity \citep{ny95a,emn97,g15}, 
therefore we assumed mass loss in the continuity equation (\ref{ce1.eq}), 
which defined as \citep{BB99},
\begin{equation}
 \dot{M}_{\rm in}=-4\pi r^2\rhe \vre {\rm cos\theta_{\rm e}}=\dot{M}_{\rm b}\left(\frac{r}{r_{\rm b}}\right)^s, 
 \label{mloss.eq}
\end{equation}
where $s$ is exponent and called as the mass-loss parameter, $r_{\rm b}$ is radial distance from the BHs  
when the disk started outflows 
from the equatorial plane and other quantities, $\rhe, \vre$ and $\theta_{\rm e}$ have denoted in appendix (\ref{sec:Feqneq}). 
According to the simulation paper by \cite{omnm05}, $s$ is not a constant in the disk but 
average value has estimated around $1$.
Since there are limitations in the analytical approach, therefore, 
we assumed `$s$' as a parameter and a constant for a particular solution.
$\dot{M}_{\rm b}$ is the mass accretion rate at radius $r_{\rm b}$. Here, 
$\dot{M}_{\rm b}=\dot{m}\dot{M}_{\rm Edd}$ and $\dot{m}$ is dimensionless mass accretion rate. 
$\dot{M}_{\rm Edd}=1.4\times 10^{17}(\mbh/\sun)(2G/c^{3})$ is the Eddington mass accretion rate in the geometrical unit.
Here, $s=0$, corresponds to constant accretion rate means no mass loss from the disk. Since we want to study the outflows, 
therefore `$s$' should be greater than zero. 
{Now the equation (\ref{mloss.eq}) after differentiation can be written as,
\begin{equation}
 \frac{2}{r}+\frac{1}{\rhe}\frac{d\rhe}{dr}+\frac{1}{\vre}\frac{d\vre}{dr}=\frac{s}{r}.
 \label{mloss1.eq}
\end{equation}
Since we assumed that the radial components of flow variables and its derivatives are same for all values of polar angle 
at or above the equatorial plane for a particular radius. 
Therefore,}
the equation (\ref{ce2.eq}) with the help of the equation (\ref{mloss1.eq}) becomes,
\begin{equation}
 \vr s+
\vt\left[\frac{1}{\rht}\frac{d\rht}{d\theta}+\frac{1}{\vtt}\frac{d \vtt}{d\theta}+{\rm cot\theta} \right]=0.
\label{ce3.eq}
\end{equation}
On solving\deleted{full model fluid equations (\ref{rc2.eq}-\ref{eg2.eq}) with equation (\ref{ce3.eq}), there are still many intricacies, so 
we made one more simplification on $\theta=\pi/2$ (the equatorial plane), 
\ie we assumed double derivative of azimuthal velocity is vanishing {\bf as one of possibility since number of unknown problem}}
{the fluid equations (\ref{rc2.eq}-\ref{eg2.eq}, \ref{ce3.eq}) with (\ref{bc.eq}) at $\theta=\pi/2$, 
we still need one more boundary condition in order to get flow variables. 
So we assumed $d^2\vp(\pi/2)/d\theta^2=0$ from following as the equation (\ref{bc.eq})}.  
Thus, the polar flow variables on $\theta=\pi/2=90$ are
estimated from equations (\ref{rc2.eq}-\ref{eg2.eq}, \ref{ce3.eq}) with using equation (\ref{bc.eq}) and 
after some simplifications, we get,
\begin{equation}
 a_{\rm e}\vrt^2(90)+b_{\rm e}\vrt(90)-F_{\rm r}=0, ~~
 \Thth(90)=\frac{x_0}{x_2}\vrt(90),~~ \vpt(90)=\sqrt{\frac{x_3}{x_4}\vrt(90)}~~\mbox{and}~~
 {\frac{d\vtt(90)}{d\theta}=-\frac{s\vr}{\vtr}},
 \label{bcv.eq}
\end{equation}
where  
$a_{\rm e}=\vrr{d\vrr}/{dr}$, $b_{\rm e}=-{\vpr^2x_3}/{(rx_4)}+{x_0x_1}/{x_2}$, ~$x_0=\vrr{d\lambda}/{dr}$, 
~$x_1=({d\Thr}/{dr}+({\Thr}/{\rh}){d\rh}/{dr})/(\tt\beta)$, ~$x_2=x_0+{\alpha_2\Thr\vpr}/({\omegk\tt\beta r})$, 
~$x_3=\vrr(N_{\rm eff}{d\Thr}/{dr}-({\Thr}/{\rh}){d\rh}/{dr})$, 
~$x_4=f{\alpha_1\Thr}({d\vpr}/{dr}-{\vpr}/{r})^2/{\omegk}$.
Here, $\vrr=\vre$, $\vpr=\vpe$, $\Thr=\Thre$, $\rh=\rhe$ and corresponding radial derivatives, 
$d\vrr/dr=d\vre/dr$, $d\vpr/dr=(d\lmda/dr-\vpe)/r$, $d\Thr/dr=d\Thre/dr$, $d\rh/dr=d\rhe/dr$ 
calculated from transonic ADAF solutions \citep{nkh97,lgy99} on the equatorial plane from equations (\ref{dt1.eq}-\ref{dl.eq}). 
Here, subscript `$e$' denotes values of the flow variables on the equatorial plane.
In next section we will 
discuss solution procedure to find critical points and ADAF solutions. 
\section{Solution Procedure}\label{sec:solnproc}
Since the BH accretion is necessarily transonic because of the nature of gravity around central objects. 
Therefore, we first define and find out the critical point of the accretion flow in following subsections.
\subsection{Critical point conditions}\label{subsec:accncp}
The critical point is a point of discontinuity of differential equation and mathematical is defined as $0/0$ form. So,
the critical point conditions are obtained from the equation (\ref{dv.eq}),
\begin{equation}
 {\cal N}=0~\Longrightarrow~\frac{(\vpe^2)_{\rm c}}{r_{\rm c}}+F_{r\rm c}+2\frac{(a_{\rm se}^2)_{\rm c}}{r_{\rm c}}+
 \frac{f}{N_{\rm eq}}(\Lambda_{\rm e}^+)_{\rm c}=0
 \label{cn.eq}
\end{equation}
and
\begin{equation}
 {\cal D}=0~\Longrightarrow~(\vre^2)_{\rm c}-(a_{\rm se}^2)_{\rm c}=0.
 \label{cd.eq}
\end{equation}
Here subscript `${\rm c}$' denotes the flow quantities at the critical point and 
the radial velocity gradient at critical points obtained by l$'$Hospital rule. 
We found critical points by satisfying equations (\ref{cn.eq}-\ref{cd.eq}) together, 
with the help of integration of equations (\ref{dt1.eq}-\ref{dl.eq}), 
for given set of parameters (${E},~\lambda_0,~\gamma,~\alpha_1$ and $\beta$), detail explanations in appendix (\ref{sec:solodes}). 
We integrated the differential equations (\ref{dt1.eq}-\ref{dl.eq}) from horizon to outward 
with the help of equation (\ref{engc1.eq}). For this, we used a very nice technique for calculation of 
asymptotic flow variables very close to the horizon,
which is describing in next subsection.
\subsection{Method to find asymptotic flow variables}
\label{subsec:methcp}
For hunting of the critical point location, we used a methodology as described in many papers \citep{bdl08,kc13,kc14,kcm14,ck16}.
Using Frobenius expansion for calculation of asymptotic value of $\lmda$ for the differential equation (\ref{dl.eq}), 
the expression is
\begin{equation}
 \lmda=\lambda_0+\zeta(r-\rs)^\Lambda, ~~~r\rightarrow\rs,
 \label{frob.eq}
\end{equation}
where $\zeta$ and $\Lambda$ are constants and to be determined by equation (\ref{dl.eq}) with using the equation (\ref{frob.eq}), we get, 
\begin{equation}
 \lim_{r\to\rs}\frac{d\lmda}{dr}=\lim_{r\to\rs}\left[\frac{2\lmda}{r}-
 \frac{\geff\vre\omegk\zeta(r-\rs)^\Lambda}{\alpha_1a_{\rm s}^2}\right],
 \label{frob1.eq}
\end{equation}
Here, we assumed $\vre=\delta v_{\rm ff}$ for limit $r\rightarrow\rs$ and $(d\lmda/dr)|_{\rs}=0$.
Where, $v_{\rm ff}=1/(r-\rs)^{1/2}$ is free-fall velocity and $\delta<1$. 
The value of $\delta$ will be obtained by iterations with satisfying the conditions (\ref{cn.eq}, \ref{cd.eq}). 
With using expressions of $\vre$ and $\omegk$
in the equation (\ref{frob1.eq}) then above equation can be written as,
\begin{equation}
 \lim_{r\to\rs}{\frac{\geff\delta\zeta(r-\rs)^\Lambda}{\alpha_1a_{\rm se}^2\sqrt{2r}(r-\rs)^{3/2}}}=\frac{2\lambda_0}{\rs}
\end{equation}
When eliminating all $(r-\rs)$ terms from the above equation then we require $\Lambda=3/2$. 
So, we get, $\zeta=2\sqrt{2}\alpha_1\lambda_0a_{\rm se}^2/(\geff\delta)$.
For a choice of $\delta$ value, we obtained a value of $\zeta$ then we calculated flow variables very close to the horizon, 
say $r=\rin=1.001$. {Now we can obtain values of $\lmda, \vre$ and $\Thre$ at $\rin$ with the help of equations (\ref{frob.eq}) 
and (\ref{engc1.eq}) then we can integrate outward fluid equations (\ref{dt1.eq}-\ref{dl.eq}) from $\rin$. 
The detail method for finding critical points (CP) and disk structure are described in the appendix (\ref{sec:solodes}).}
\deleted{When we combined equations (\ref{frob.eq}) and (\ref{engc1.eq}) with the value of $\Lambda$ and expression of $\zeta$ at $\rin$. 
Thus we got a polynomial in $\ase$ or $\Thre$. 
Now, supplying the parameters ${\cal E}$ (or $E$), $\lambda_0$, $\alpha_1$, $\gamma$ and $\beta$ then 
we solved the polynomial for $\Thre$ at $\rin$ for first choice $\delta=1$. 
Once $\Thre$ obtained, other quantities $\vre$ and $\lmda$ easily get with the help of the $v_{\rm ff}$ and equation (\ref{frob.eq}). 
We now can integrate differential equations (\ref{dt1.eq}-\ref{dl.eq}) outward from $\rin$ by using $\Thre$, $\vre$ and $\lmda$ and 
simultaneously, checked the sonic point equations (\ref{cn.eq}-\ref{cd.eq}). 
If sonic conditions are not satisfied then we reduced the value of $\delta$ and 
repeat the whole procedure again from the solving polynomial for $\Thre$ to checking sonic point conditions. 
This solution procedure repeated till satisfy sonic conditions. 
When ensuring it then we obtained critical point location ($r_{\rm c}$) for given flow parameters. 
Once $r_{\rm c}$ obtained, we integrated equations (\ref{dt1.eq}-\ref{dl.eq}) outward along the radial direction 
for a given $\lambda_0$ with other disk parameters. 
Then we investigated outer boundaries of ADAF solution \citep{nkh97,lgy99} by changing $\lambda_0$ with repeating whole above procedure.
Once $\lambda_0$ obtained for the ADAF solution, we simultaneously integrated ODEs ( the radial equations \ref{dt1.eq}-\ref{dl.eq} and the 
polar equations \ref{rc2.eq}-\ref{eg2.eq} with \ref{ce3.eq}) along the radial direction (inward and outward) 
 and along the polar direction with using obtained radial flow variables and 
symmetric boundary conditions on the equatorial plane, we got complete 2D disk structure of the fluid flows.} 
\section{Numerical results}\label{sec:nresult}
We solved analytically 2D fluid equations with assuming explicitly radial fluid equations on the equatorial plane and 
first integrated along the radial direction, say at $r$ then immediately at same $r$, we solved along the polar direction 
from the equatorial plane (details in the appendix \ref{sec:solodes}).
Since there are many analytical studies on the 2D disk structure with using ADAF self-similar assumptions 
on the equatorial plane \citep{ny95a,xc97,BB99,XW05,JW11}. 
Therefore, we investigated the ADAF solutions on the equatorial plane for the calculations of the polar flow variables.  
So we would first represent the ADAF solutions in coming subsection and later next subsection with complete inflow-outflow solutions. 
In present work, we used both extreme values of $\gamma$ or $\geff$, one $\gamma=5/3$, 
where $\geff$ depends on $\beta$, which may change the disk flow variables and structure with changing $\beta$.
And other $\gamma=4/3$, where $\geff=\gamma$ for any value of $\beta$.  
Here the mass inflow density and pressure of the gas have been calculated 
with the mass accretion rate $\dot{m}=0.1$ and $\mbh=10\sun$ for all the solutions of this paper. 
\subsection{ADAFs solutions}\label{subsec:adaf}
We used flow parameters to find the transonic accretion solutions on the equatorial plane, which are ${E}, \lambda_0, \gamma, \beta$ and $\alpha_1$. 
\begin{figure}[ht!]
\plotone{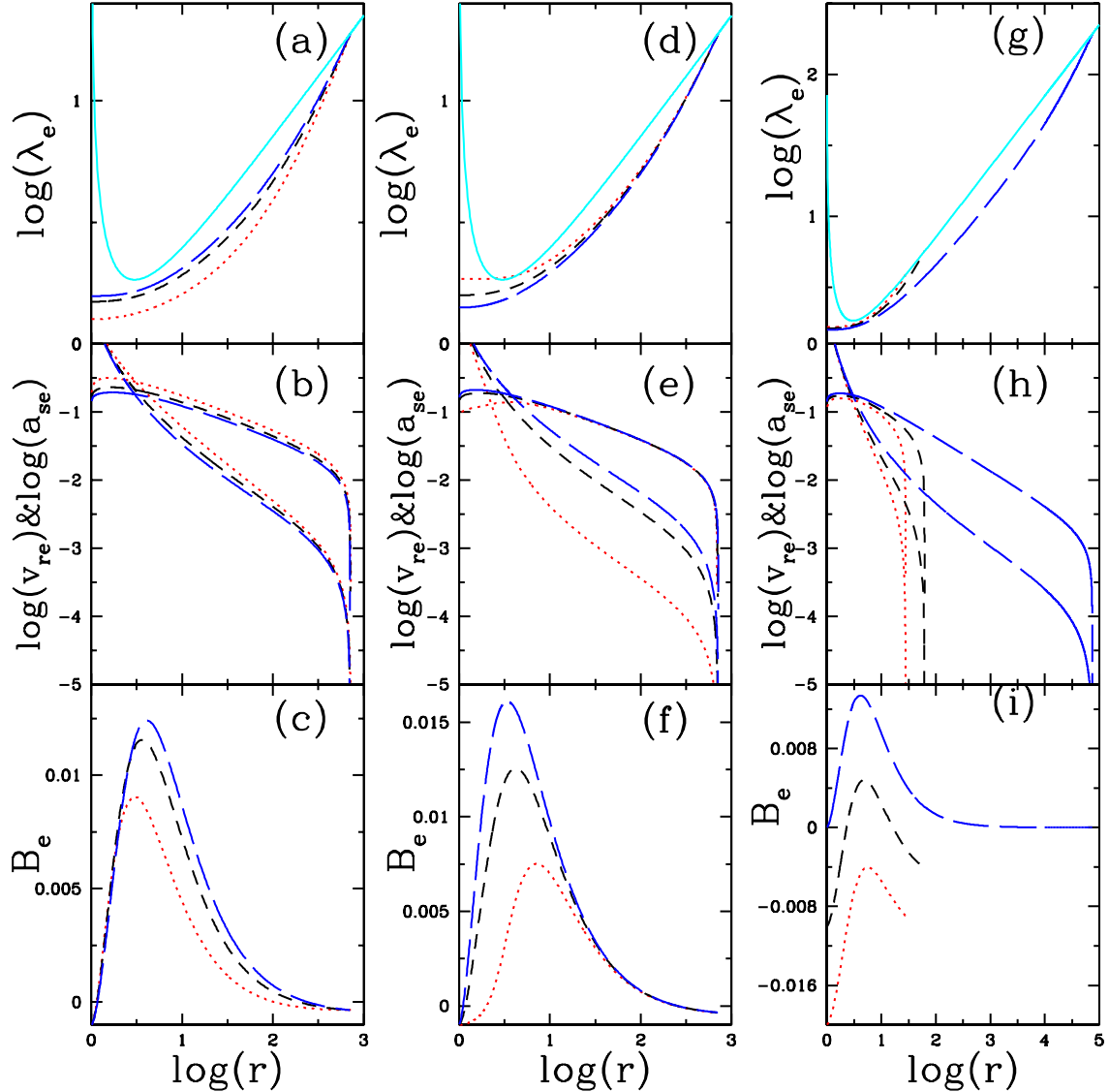}
\caption{Variations of the radial flow variables with radial distance, $log(r)$. Panels are showing variation of $log(\lmda)$ (a, d, g), $log(\vre~ \& ~\ase)$ (b, e, h) 
and $\be$ (c, f, i).
The panels (a-c) are plotted for parameters, ${E}=-0.001, \alpha_1=0.1, \gamma=5/3$ with different $\beta=1$ (dotted red), $0.5$ (dashed black) 
and $0.1$ (long-dashed blue). The panels (d-f) are plotted for energy parameter ${E}=-0.001$ with
different $\alpha_1=0.01$ (dotted, red), $0.1$ (dashed, black) and $0.2$ (long-dashed, blue).
The panels (g-i) are plotted for viscosity parameter $\alpha_1=0.1$ with different ${E}=-0.02$ (dotted, red), $-0.01$ (dashed, black) 
and $-0.00001$ (long-dashed, blue). 
The second and third columns are plotted for same $\gamma=\geff=4/3$ and $\beta=1$.
The solid curve (cyan color) represented the Keplerian angular momentum distribution in panels (a, d, g).
\label{fig2}}
\end{figure}
In Figure (\ref{fig2}), we represented typical ADAF solutions as previously shown by \cite{nkh97,lgy99}. 
Here, which are plotted with different values of $\beta$ in a first column, 
values of viscosity parameter ($\alpha_1$) in a second column, and 
values of the grand specific energy (${E}$) of the flow in a third column, which changes $r_{\rm t}$. 
Here $r_{\rm t}$ is outer boundary of the ADAF or assumed transition radius from the Keplerian to the sub-Keplerian flows of 
the two zone configuration of the disk.
The distribution of specific angular momentum ($\lmda$), bulk velocity ($\vre$) with sound speed 
($\ase$) and the Bernoulli parameter $\be$ 
are plotted in panels \ref{fig2}(a, d, \& g), panels \ref{fig2}(b, e, \& h) and panels \ref{fig2}(c, f, \& i), respectively. 
In panel (\ref{fig2}a), the value of $\lmda$ is lowest for $\beta=1$ and increases with decreasing $\beta$,  
when keeping other parameters are fixed. 
Since $\lmda$ is lower for $\beta=1$, therefore $\vre$ and $\ase$ are higher in panel (\ref{fig2}b). 
The $\be$ is lower for higher $\beta$ in panel (\ref{fig2}c), since the $\lmda$ is low, which is not compensated by high values of $\vre$ and $\ase$.
In the second column of the Figure (\ref{fig2}), we changed $\alpha_1$ and kept other parameters same. 
In panel (\ref{fig2}d), values of $\lmda$ is higher for $\alpha_1=0.01$. 
Since angular momentum transported less for lower viscosity when $r_{\rm t}$ or ${E}$ is same. 
Therefore $\vre$ and $\ase$ are lower in panel (\ref{fig2}e) and $\be$ is also low in panel (\ref{fig2}f), which is not compensated by higher $\lmda$. 
The third column of the Figure (\ref{fig2}), we plotted curves with different ${E}$ means changing $r_{\rm t}$ and 
keeping other parameters fixed. 
Here, the distribution of $\lmda$ are almost same close to the BH and higher 
when approaching to $r_{\rm t}$ for lower ${E}$ in panel (\ref{fig2}g). 
As the expected variation of $\vre$ and $\ase$ are low for corresponding high $\lmda$ around $r_{\rm t}$ in panel (\ref{fig2}h). 
The $\be$ is lower for lower ${E}$ in panel (\ref{fig2}i), so this may indicates for shorter $r_{\rm t}$, 
the possibility of the outflows may weak. 
Here, $r_{\rm t}$ for different $E=-0.02, -0.01, 0.001$ and $-0.00001$ are $28, 62, 705$ and $74240$, respectively. 
{We get same power law scaling with the radius for $\ase$ and $\lmda$ as in the \cite{nkh97} and 
almost independent of $r_{\rm t}$ but 
the scaling for $\vre$ is changing significantly with $r_{\rm t}$ (or $E$) as in the Figure (\ref{fig2}h), 
\eg $\vre\propto r^{-1}$ for $r_{\rm t}\approx 700$, $\vre\propto r^{-0.7}$ 
for $r_{\rm t}\approx 70000$ and $\vre\propto r^{-0.5}$ for $r_{\rm t}\approx 2\times 10^6$ (as seen in \cite{nkh97}). 
Interestingly, these scaling rules are also same for the total $\vr, a_{\rm s}$ and $\lambda$ ($=r\vp$) on the equatorial plane, 
when calculating the 2D structures.}
Moreover, all the sub-Keplerian solutions have $\Theta>1$ in the vicinity of the BHs but around $r_{\rm t}$ have $\Theta\ll 1$ and 
mass density will be higher at $r_{\rm t}$ since $\vre\sim 0$, 
so before transition radius, we believe that flow was Keplerian.
Since the sub-Keplerian flows are showing positive $\be$ in the intermediate values of $r$, 
therefore may give rise outflows \citep{ny95a}. 
Therefore we used these sub-Keplerian hot flows for the generation of the outflows and 
investigated the 2D disk structures as presented in next subsection.
\subsection{Inflow-outflow solutions}\label{subsec:ioflw}
Here outflow solutions above the equatorial plane in $\theta-$directions are calculated only up to a sonic surface 
when outflow Mach number ($M=|\vec{v}|/a_{\rm s}$) becomes equal to one, 
where $\vec{v}=\vec{\vr}+\vec{\vt}$ is the total velocity of fluid and $a_{\rm s}^2={\geff P/\rho}$ is a sound speed.
Since the sonic surface rises a kind of discontinuity in the analytical integration of differential equations, 
therefore, integrations are invalid after the sonic surface without taking any proper methodology to solve the discontinuity. 
Therefore the fate of these outflows after the sonic surface is unknown in this study but we can predict that  
these disks may have strong outflows on the basis of transonic nature of the outflows and $M$ will reach very large after 
crossing the sonic surface. 
Here we used parameters $s$ and $\alpha_2$ are non-zero when ODEs integrated along the polar directions.
All the 2D disk figures with velocity vectors and density contours are plotted up to the radius size ($r_{\rm b}$), 
where outflows are started to generate from the disk. 
{Here we are redefined a few flow variables in their physical units, \eg the flow density $\bar{\rho}=\rho f_{\rm\rho}$, 
the gas pressure $\bar{p}_{\rm g}=p_{\rm g}f_{\rm p}$ and flow temperature $T=\Theta f_{\rm T}$. 
Here, $f_{\rm\rho}=7.75\times 10^{16}(\dot{m}/m)g~cm^{-3}$, $f_{\rm p}=6.98\times 10^{37}(\dot{m}/m)g~cm~s^{-2}$, where 
$\dot{m}$ and $m$ are accretion rate in unit of the Eddington accretion rate and mass of the BH in unit of the solar mass, respectively 
and $f_{\rm T}=5.93\times 10^9K$.}\\ 
\begin{figure*}
\gridline{\fig{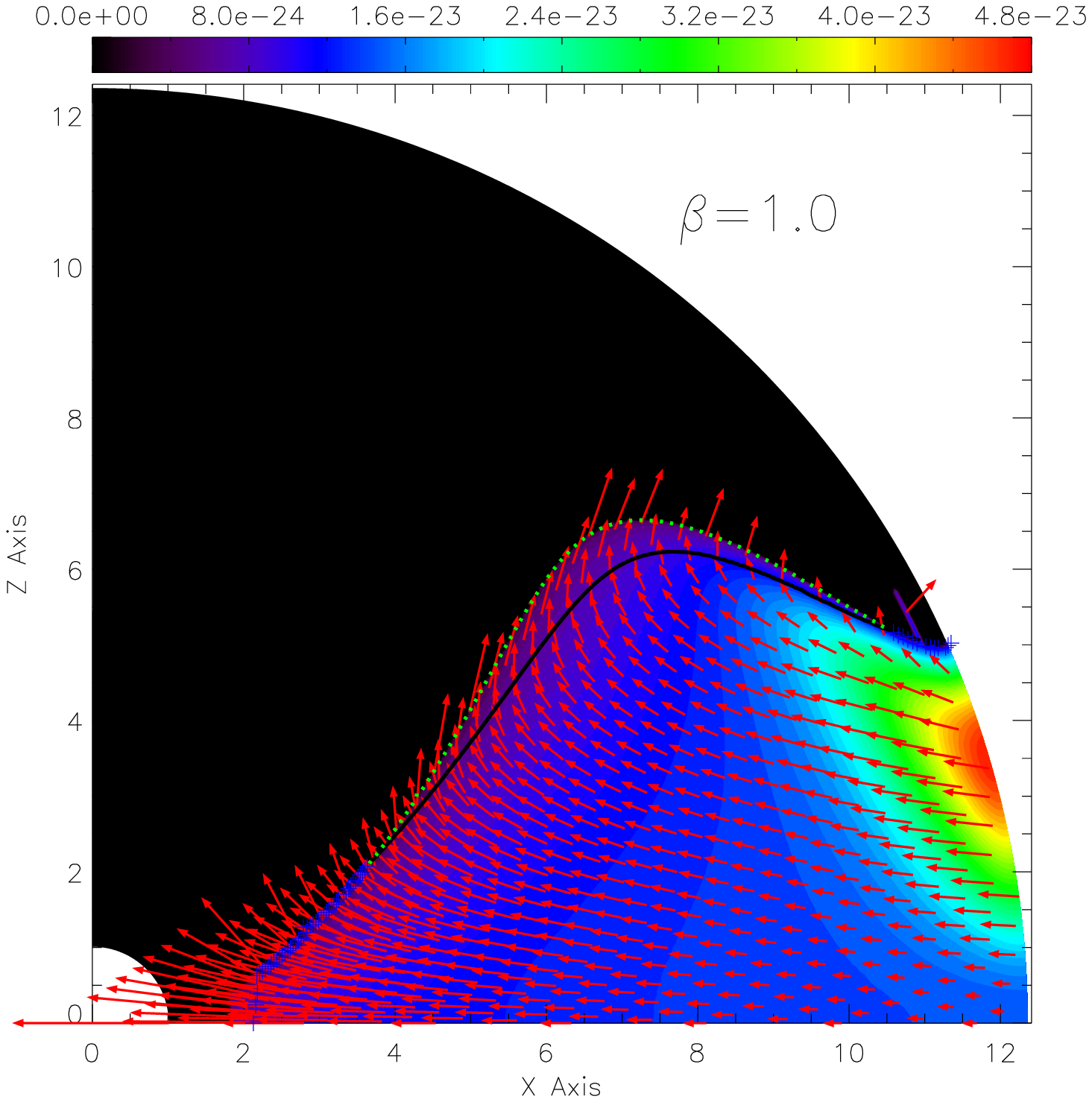}{0.3\textwidth}{(a)}
         \fig{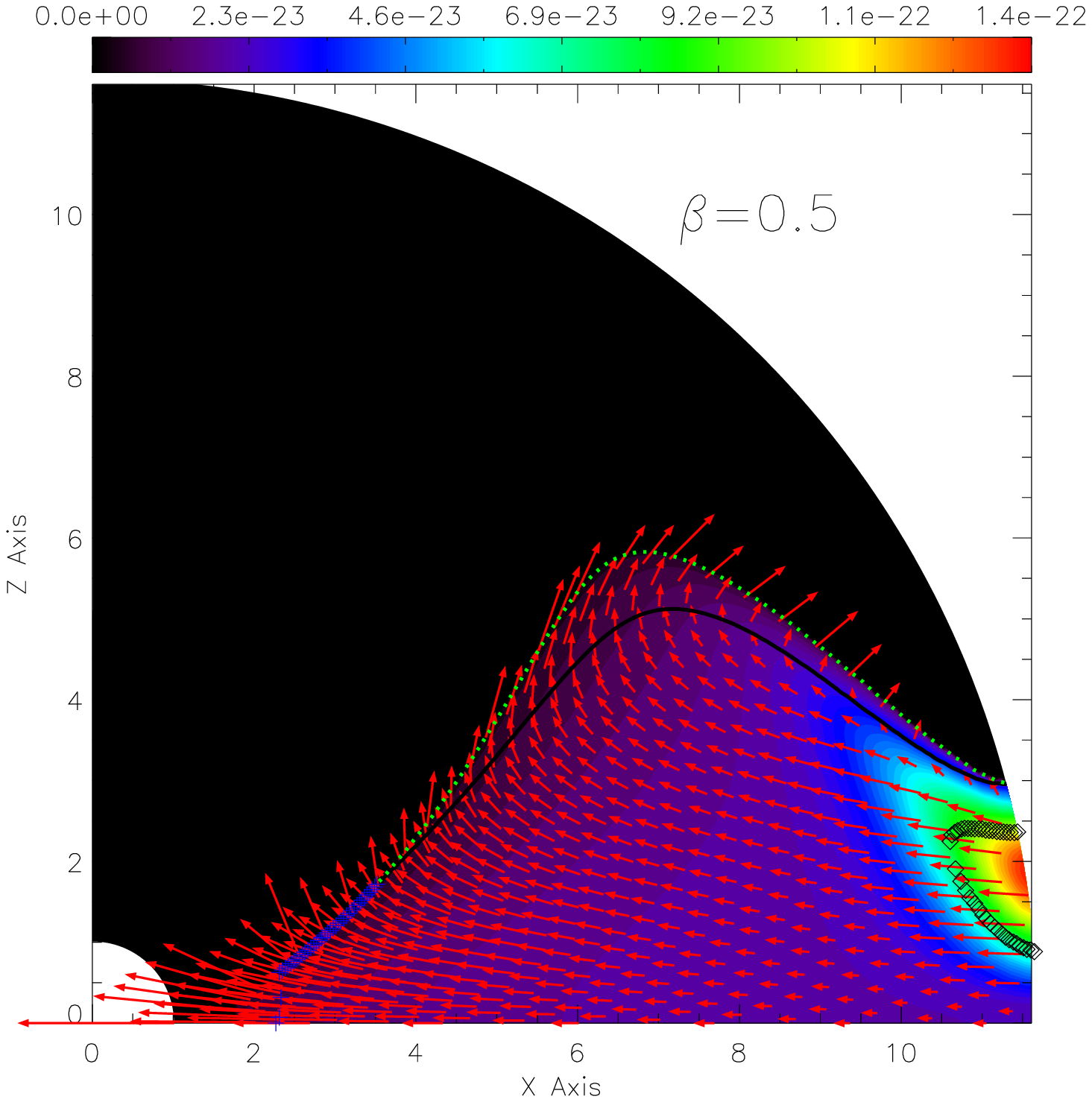}{0.3\textwidth}{(b)}
         \fig{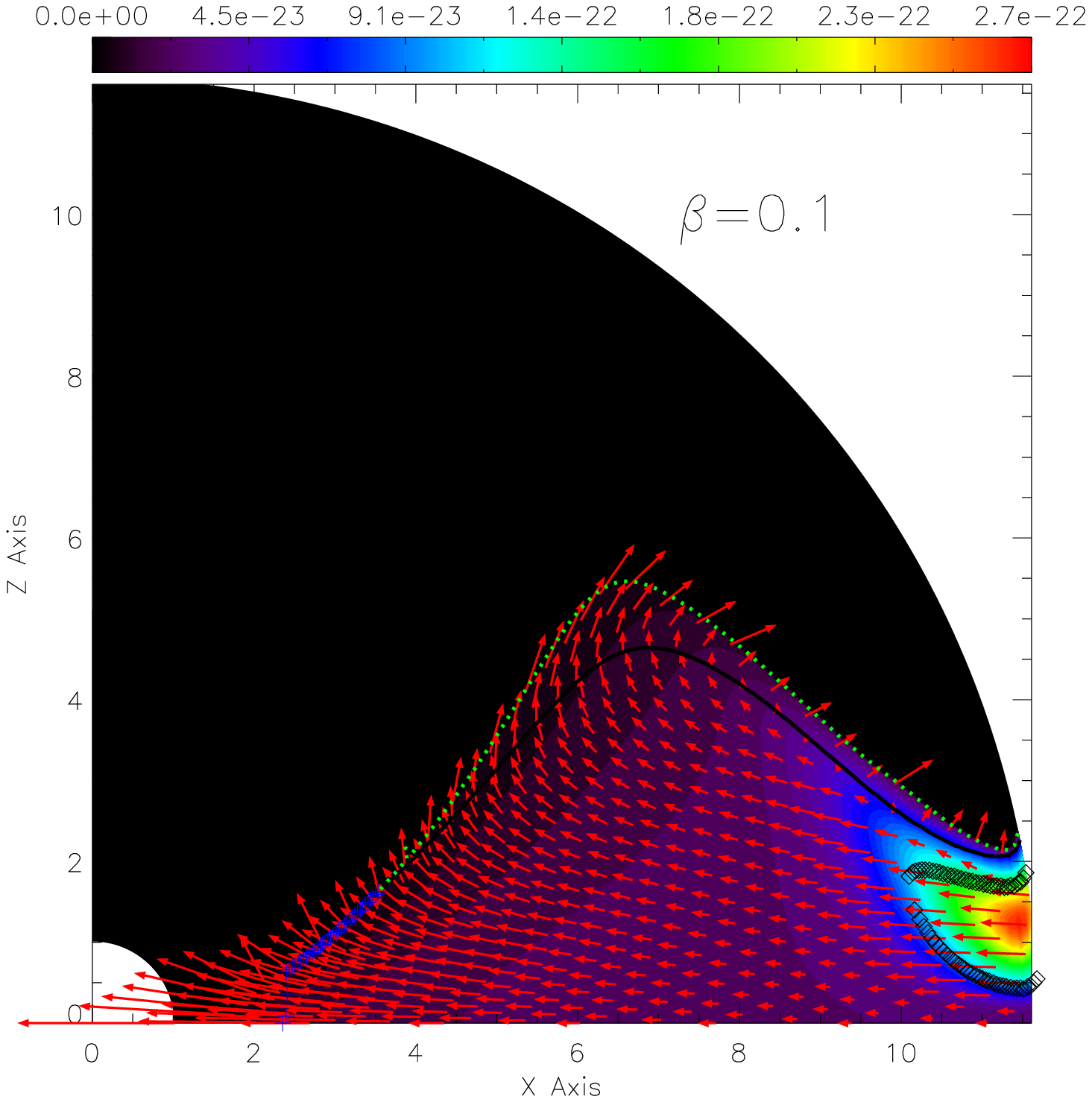}{0.3\textwidth}{(c)}
         }
\gridline{\fig{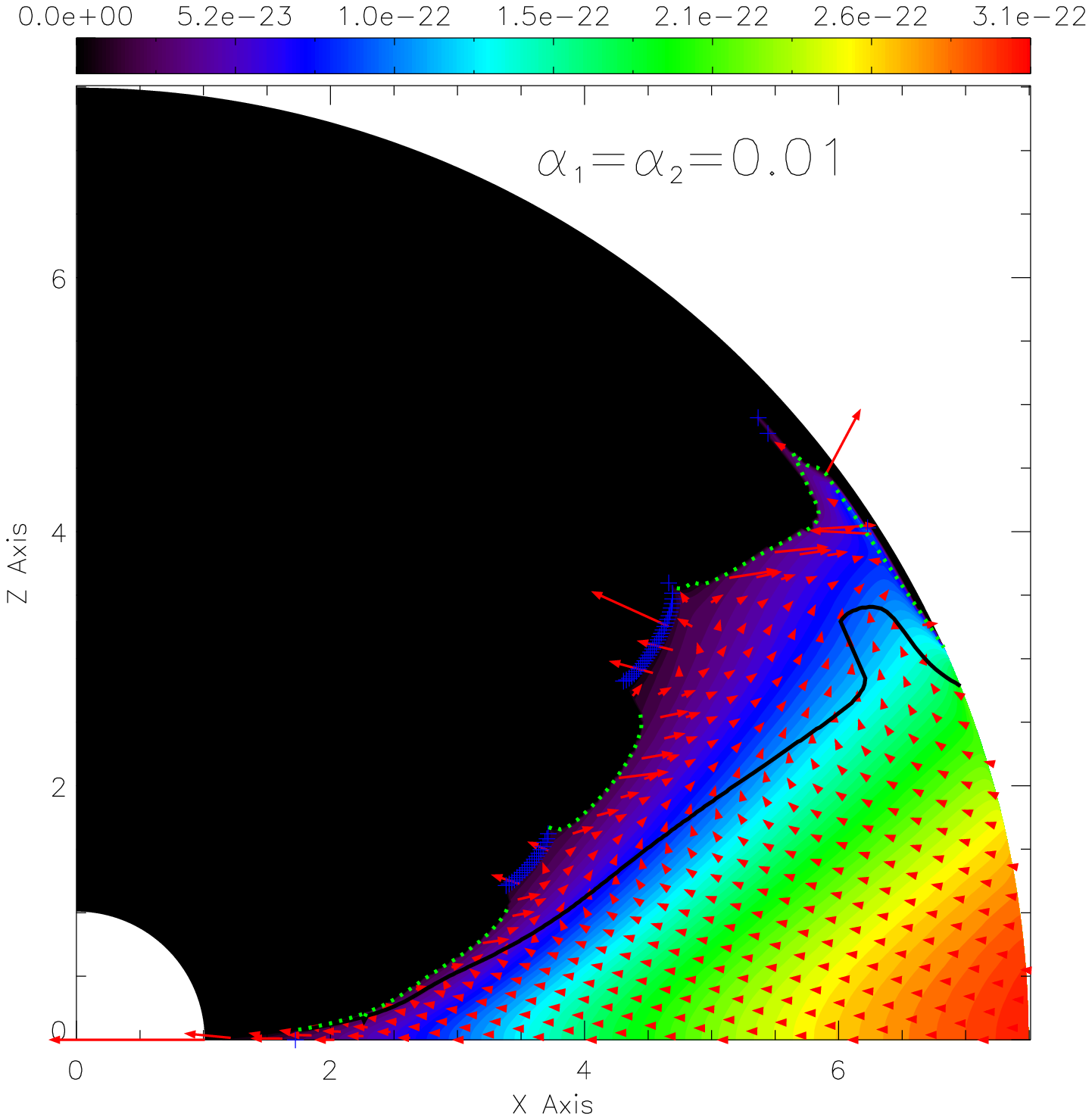}{0.3\textwidth}{(d)}
          \fig{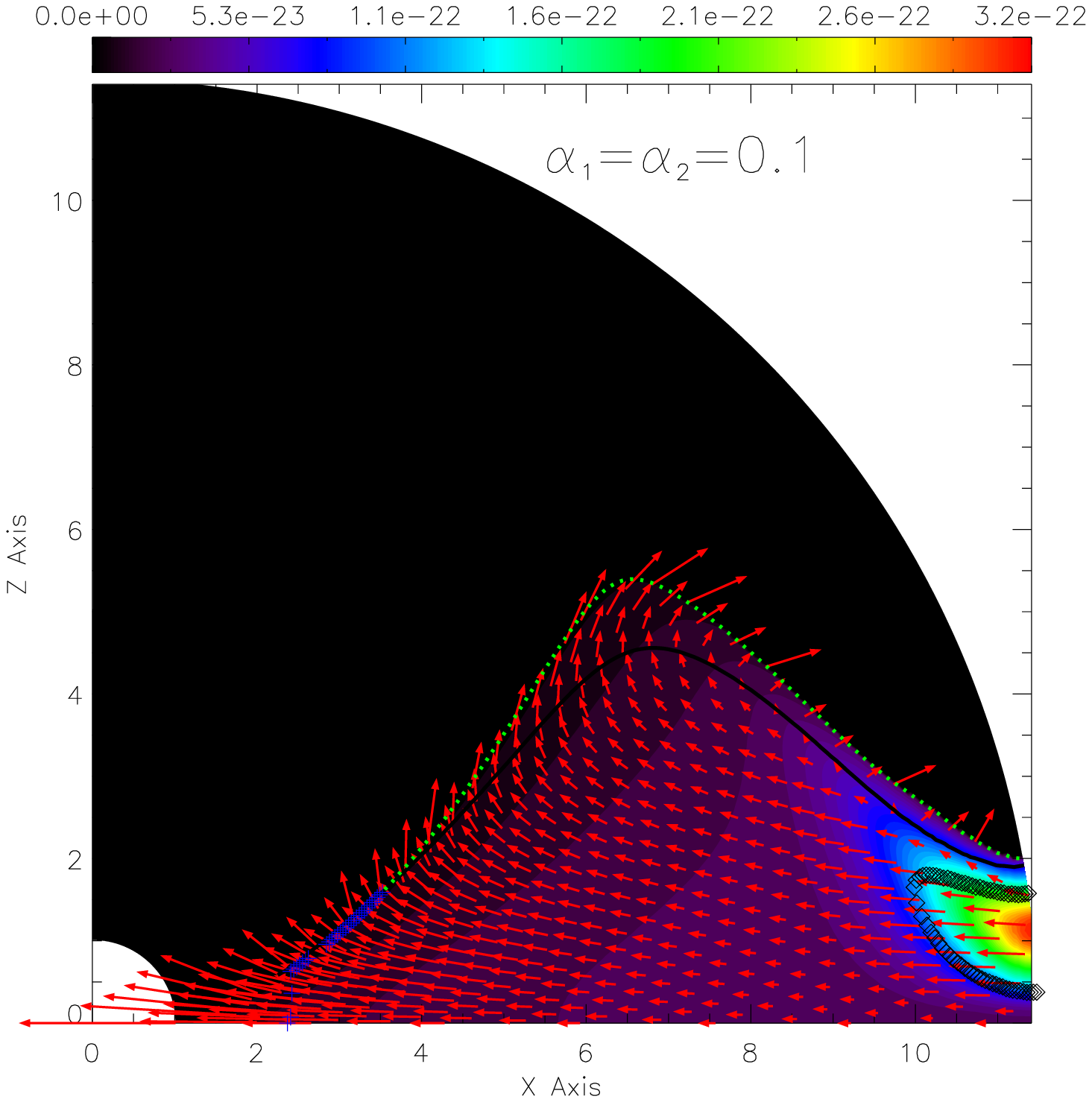}{0.3\textwidth}{(e)}
          \fig{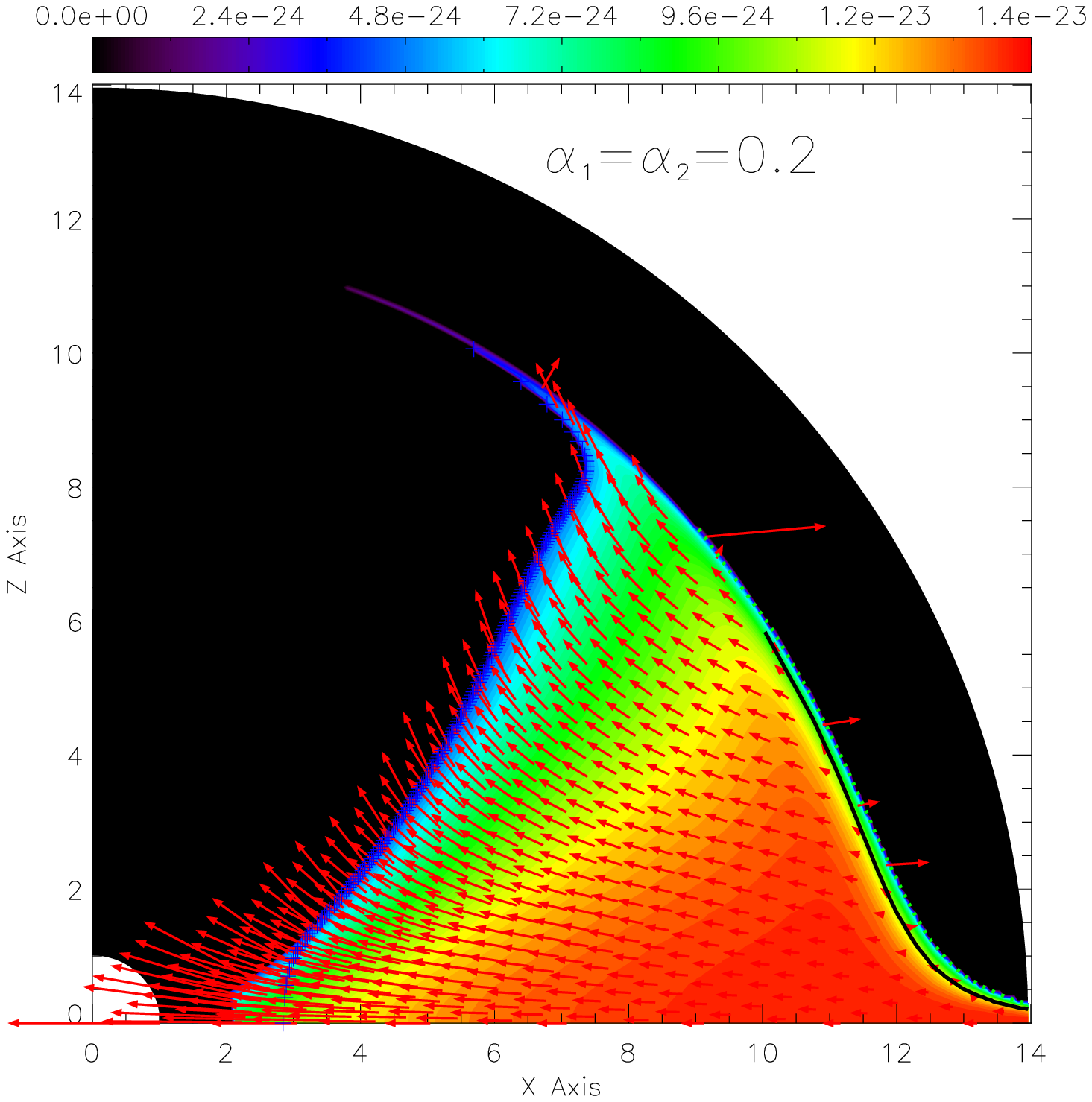}{0.3\textwidth}{(f)}
          }
\gridline{\fig{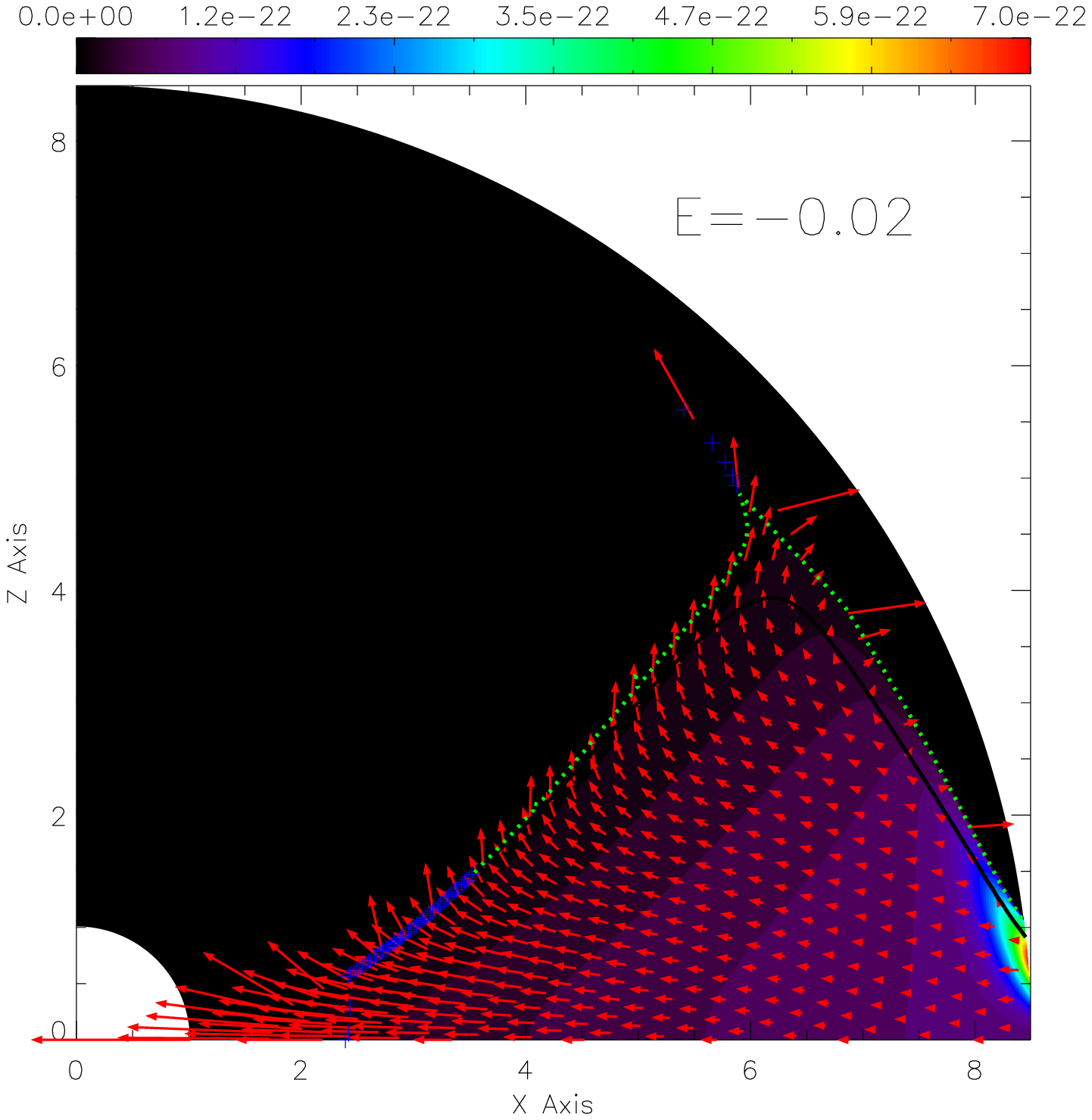}{0.3\textwidth}{(g)}
          \fig{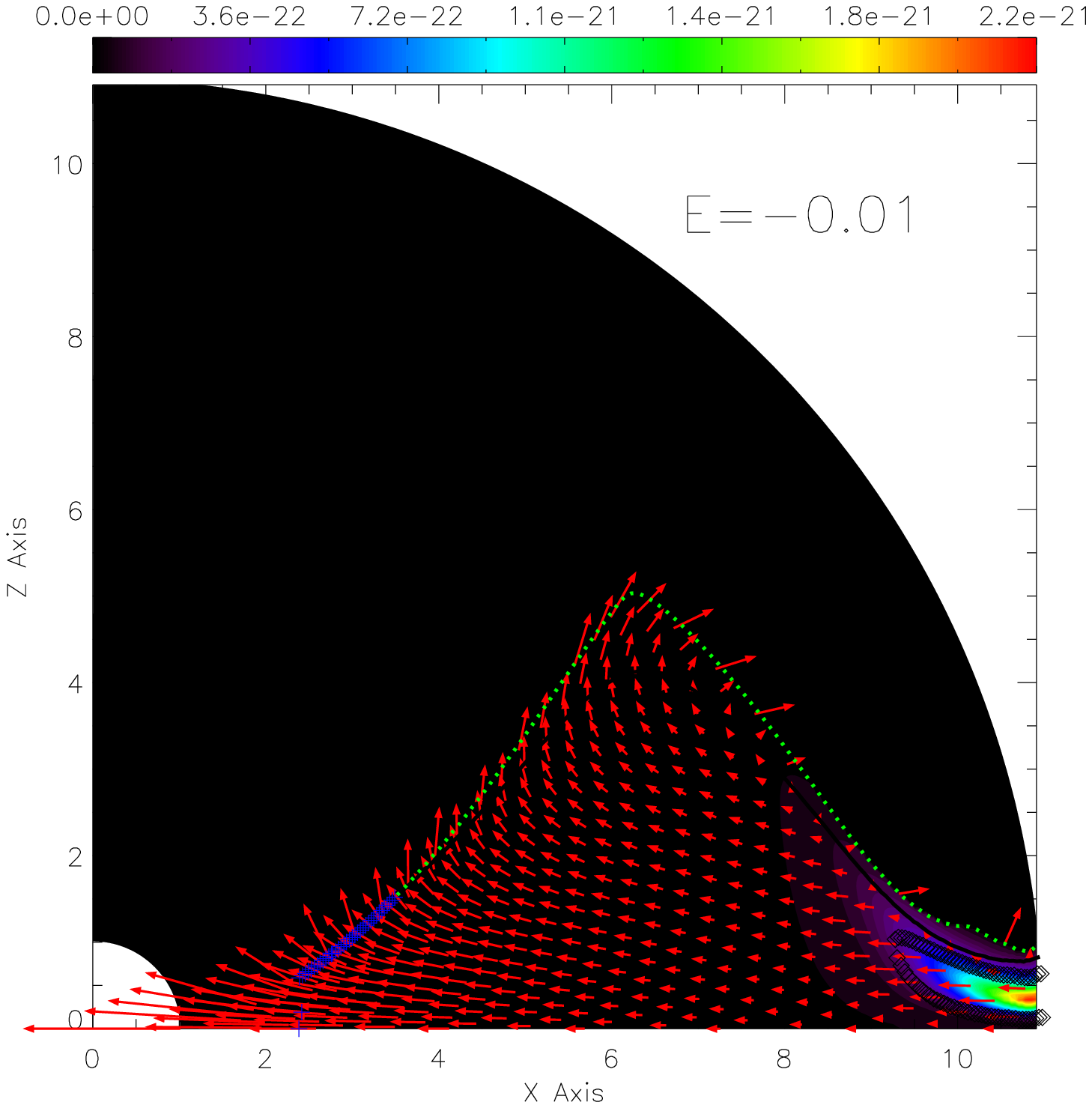}{0.3\textwidth}{(h)}
          \fig{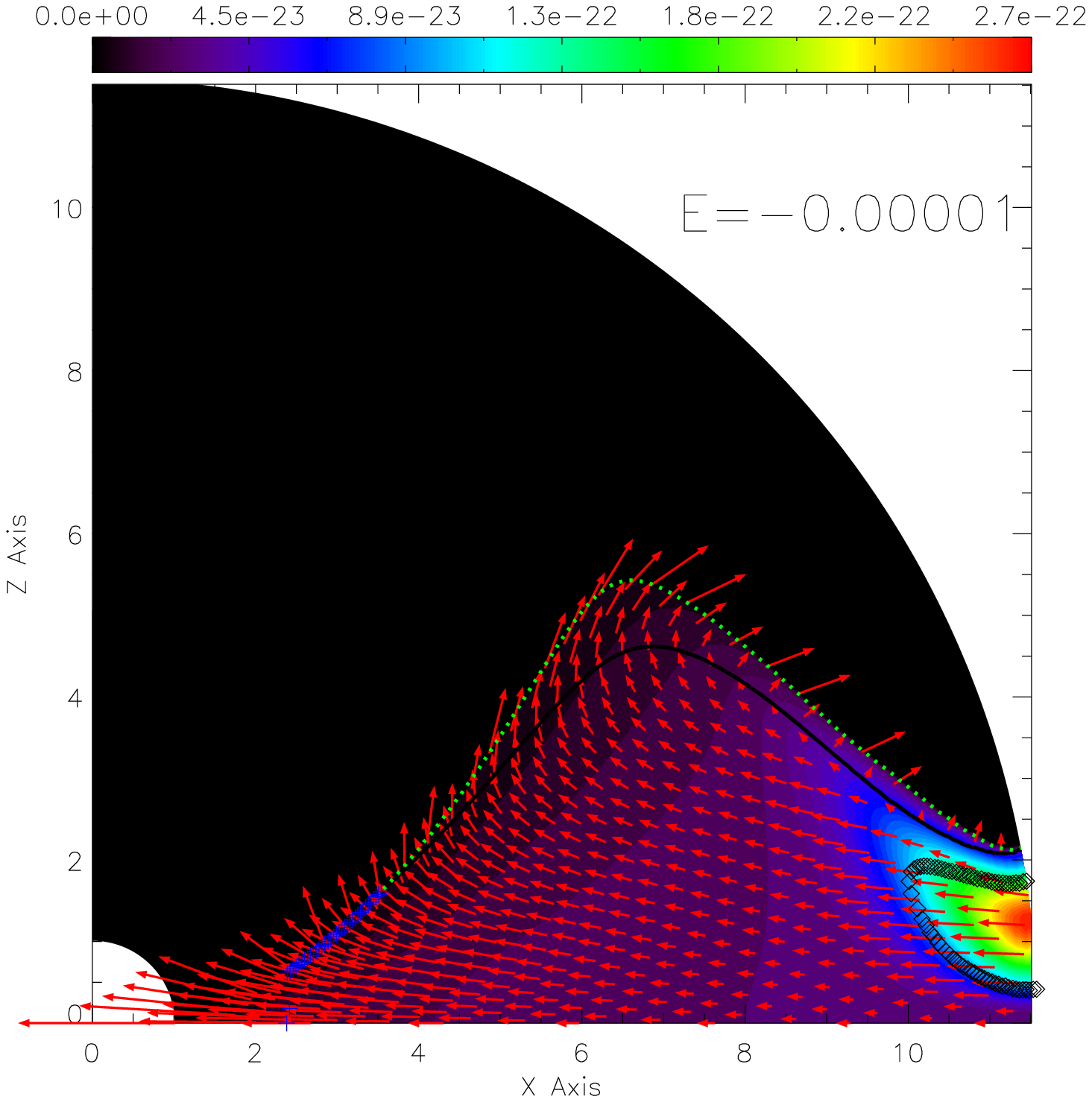}{0.3\textwidth}{(i)}
          }
\caption{The 2D disk structure represented with inflow density contour and velocity vector field, which is representing velocity direction of 
$\vec{v}=\vec{\vr}+\vec{\vt}$ (or $\vec{v}_{\rm j}$) with a magnitude of $|\vec{v}|$ (or $|\vec{v}_{\rm j}|$). 
A first, second and third rows are plotted for different $\beta$, viscosity parameters ($\alpha_1 ~\& ~\alpha_2$) and ${E}$, respectively. 
All the panels are plotted with corresponding same disk parameters of the Figure (\ref{fig2}) 
with $s=1$ and $\alpha_2=\alpha_1$. The first row has $\gamma=5/3$ and other two rows have $\gamma=\geff=4/3$. 
Here solid black, dotted green, `+' symbol blue and open square symbol black curves are representing disk surface, 
outflow sonic surface, inflow sonic surfaces and inflow supersonic region, respectively. 
Color bar at top of figures represents inflow density variation. 
Here the axes and density are having unit of $\rs$ and $f_{\rm \rho}=7.75\times 10^{14}g~cm^{-3}$, respectively.
\label{fig3}}
\end{figure*}
Figure (\ref{fig3}) is representing the 2D disk structures with velocity vector fields and density contours. 
In a first row of the Figure (\ref{fig3}) is plotted with  
different $\beta=1$ (panel \ref{fig3}a), $0.5$ (panel \ref{fig3}b) and $0.1$ (panel \ref{fig3}c), 
which are corresponding radial input solutions of the first column of the Figure (\ref{fig2}). 
Disk thickness is high for the panel (\ref{fig3}a) in the row. 
The disk thickness decreases with decreasing $\beta$ towards panel (\ref{fig3}c), which is radiation-dominated for $\beta=0.1$. 
Since the flow is accelerated more with decreasing $\beta$, therefore, the gas formed the disk surface at lower latitude and  
the radiation dominated flows may have strong outflows.
Moreover, the inflow matter becomes supersonic close to the horizon after `+' symbol blue color line 
(where $\vr<0$ and $M=1$). 
Here, the disk surface ($\vr=0$ and $M<1$) and outflow sonic surface ($\vr>0$ and $M=1$) are represented with solid black line and 
dotted green line, respectively. The disk surface (solid line) separated the inflow and outflow regions in the disk. 
The velocity vectors are mostly directed towards the BH in the inflow region and in the outflow region, they are going out.
{Here, we found that the gas or radiation pressure dominated flows are having the outflows. Which are consistent with simulations
for both the gas and radiation pressure supported flows. The case with advection-dominated and $\beta<1$ can 
resemble for high accretion rate flows with high luminous BH sources. The case with advection-dominated and $\beta\approx 1$ can 
resemble for low accretion rate with low luminous BH sources.} 
In a second row of the Figure (\ref{fig3}) is plotted with different viscosity parameters $\alpha_1=\alpha_2=0.01$ (panel \ref{fig3}d), 
$0.1$ (panel \ref{fig3}e) and $0.2$ (panel \ref{fig3}f) for the input parameters corresponding to solutions of 
second column of the Figure (\ref{fig2}). 
In panel (\ref{fig3}d), which is plotted with $\alpha_1=\alpha_2=0.01$ we get two kind of the sonic surfaces 
above the disk surface (solid black), one for the outflow, where $\vr>0$ (dotted green) and 
other for the fail outflows (`+' blue), when the solutions fail to make sonic transition in the outflow and 
matter velocity again becomes $\vr<0$. 
Corresponding to these two surface regions, the detail variations of flow variables along $\theta-$ direction 
are presented below in Figure (\ref{fig4}). 
In the second row, we increased the value of viscosity parameters and 
panel (\ref{fig3}e) is plotted with $\alpha_1=\alpha_2=0.1$ then 
we got smooth inflow and outflow surfaces. The outflow strength and region are also increased. 
If we further increased the viscosity, $\alpha_1=\alpha_2=0.2$ (panel \ref{fig3}f) then 
we got almost no outflows with 2D disk inflow structure. 
Typical behavior of the flow variables of the second row panels are presented in Figure (\ref{fig5}) with some fixed radius. 
{In these cases, the outflows are strongly depends on the viscosity and 
with changing viscosity we can get no outflows, weak and strong outflows. They may explain various states of the BHCs, 
since the viscous time and cooling time scales are changed with changing viscosity or mass accretion rate parameter \citep{ds13}. 
Both the parameters are also changed the distribution of angular momentum in the flow \citep{kc14}, 
so the flowing matter can becomes Keplerian or sub-Keplerian.}
Now in last row, we are represented 2D structure corresponding to solutions of last column of the Figure (\ref{fig2}). 
Here we found that the outflow region and strength depend on the transition radius and 
both are increased with increasing $r_{\rm t}$. 
The $r_{\rm t}$ depends on the ${E}$.
For small $r_{\rm t}$, the outflow region is small and strength is also weak due to low local energy of the flow, 
when we compared with panels (\ref{fig3}g), (\ref{fig3}h) and (\ref{fig3}i) in the row, 
which are having transition radius $r_{\rm t}=28$ (${E}=-0.02$), $62$ (${E}=-0.01$) 
and $74240$ (${E}=-0.00001$), respectively. 
If we compare with the panel (\ref{fig3}e), which has $r_{\rm t}=705$ (${E}=-0.001$) and the panel (\ref{fig3}i)  
then they give almost same disk structures because both have small flow energy difference. 
So, for the small advective disk ($r_{\rm t}<100\rs$), the outflow region and strength are much affected 
by changing the ADAF disk size 
{and accordingly the Keplerian disk size is also changed as explained by \cite{emn97}. 
However in \cite{emn97} paper the ADAF size is decreased by increasing mass accretion rate, 
which also changed the distribution of local energy of the flow 
as similarly $E$ did it in the present paper}. 

In the Figure (\ref{fig3}), the inflow matter formed the sonic surface very close to the BH ($r<4\rs$) and hence 
matter enters into the BH supersonically. 
Moreover, the inflow region away from the BH ($r>10\rs$) also becomes supersonic 
before the outflows are started to generate from the disk, 
which is marked by $\square$ (open square) black symbol in the second and 
third columns of the Figure (\ref{fig3}), except panel (f). 
But the inflow matter is always subsonic very close to the equatorial plane before inner critical point or sonic surface ($r<4\rs$). 
These supersonic regions are having locally higher density, lower $T$, high $\vr<0$ and high rising $\vt$, 
so total local velocity $\vec{v}=\vec{\vr}+\vec{\vt}$ is high therefore arrow length is large in the same regions. 
The more detail of variations of the flow variables, we will present in Figure (\ref{fig6}). 
Interestingly, the inflow matter is supersonic before making outflows. 
As the matter is moving inward, the flow variables are changed very fast and 
the flow becomes subsonic. 
This supersonic to subsonic transition of the inflow matter along the radial direction may give hint 
for the possibility of occurrence of shock transition 
in the flow, although this transition is not much sharp as accretion shocks studied in the literature 
\citep{f87,c89,bdl08,kscc13,kcm14,ck16,lckhr16,kc17}. 
The outflows occurred close to the BH because the thermal pressure and rotation velocity are increasing very fast 
and the local energy (as the profile of $\be$ in the Figure \ref{fig2}c, f and i) becomes sufficient to generate bipolar outflows. 
We get disk surface with changing slope at every radius, which is 
unlike to the previous studies of 2D disk structure with self-similar assumptions \citep{JW11}, 
they got the inflow disk surface with a constant slope.  
We also investigated no outflows, outflows and failed outflows regions of the 2D flow, which are depending on the disk parameters.

The above descriptions of the Figure (\ref{fig3}) have based on the observations of the velocity fields and the outflow size. 
Now for more detail study of these figures, we plotted typical flow variables along $\theta-$direction with some fixed radius. 
\begin{figure}[ht!]
\plotone{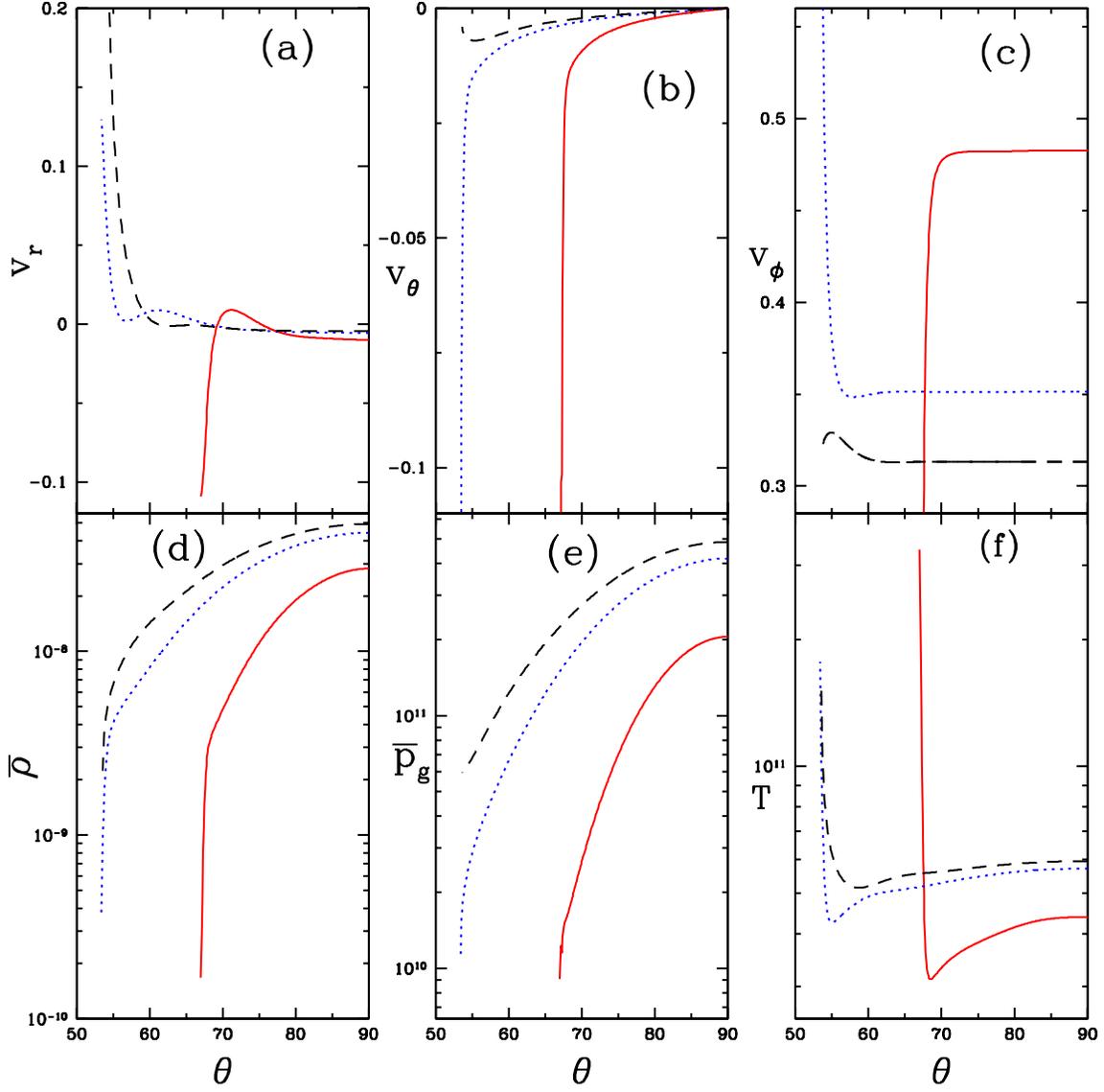}
\caption{Variations of flow variables with polar angle, $\theta$. Panels are showing variations of radial velocity, $\vr$ (a), 
evaporation velocity, $\vt$ (b),  
azimuthal velocity, $\vp$ (c), mass density, $\bar{\rho}$ (d), gas pressure, $\bar{p}_{\rm g}$ (e) and temperature, $T$ (f). 
These solutions are drawn with different radii, $r=4$ (solid, red), $6$ (dotted, blue) and $7$ (dashed, black).
All curves are drawn from the Figure (\ref{fig3}d). 
Here all the velocities, $\bar{\rho}$, $\bar{p}_{\rm g}$ and $T$ are having units of $c$, $g~cm^{-3}$, 
$g~cm^{-1}~s^{-2}$ and $K$, respectively.
\label{fig4}}
\end{figure}
The Figure (\ref{fig4}) is presented for three radii along the $\theta-$direction with same disk parameters, 
which are corresponding to the panel (\ref{fig3}d). 
These are the typical outflow solutions with low viscosity values and we found variations of the flow variables are different from the 
previous analytical studies \citep{xc97,XW05,JW11} but have some basic qualitative similarities with the outflow solutions. 
The basic properties of the outflows are the radial velocity should becomes $\vr>0$ at some `$\theta$' above the equatorial plane and 
other flow variables behavior may vary with boundary conditions on the equatorial plane.
Here we get failed outflow solutions (solid red, $r=4$), which is multi-valued solution means $\vr$ has same value at various $\theta$, 
outflow solution (dotted blue, $r=6$), which is also multi-valued and 
radially out outflow solution (dashed black, $r=7$) means $\vt\rightarrow 0$ at high latitude. 
Since these outflows are mainly driven by combinations of centrifugal force and gradient of pressures (gas or radiation) force, 
and behavior of temperature and angular velocity vary with the radial distance, therefore outflows and disk structure changed with 
the radius. If we see panel (\ref{fig4}a), initially the $\vr<0$ for inflow and at some `$\theta$' becomes zero, 
which gives the disk surface and $\vr>0$ gives outflow region. 
In panel (\ref{fig4}b), the polar velocities ($\vt<0$) are increasing with decreasing `$\theta$' but dashed black curve 
again turn back and approaches towards zero because $\vp$ is started to decreasing at high latitude (panel \ref{fig4}c).   
The $|\vt|$ is higher for lower radius solutions, this behavior maybe due to corresponding higher rotation velocity  
as in the panel (\ref{fig4}c). 
The gas density $\bar{\rho}$ (panel \ref{fig4}d) and pressure $\bar{p}_{\rm g}$ (panel \ref{fig4}e) are monotonically decreasing 
towards axis due to the expansion of gas above the equatorial plane. 
\deleted{Here $\rho$ and $p_{\rm g}$ are normalized with minimum values of $\rho$ and $p_{\rm g}$ on the equatorial plane, respectively.} 
The behavior of temperature is also not monotonic, it is decreasing and increasing toward axis in panel (\ref{fig4}f) 
{due to multi-valued nature of $\vr$. And nature of $\vr$ is mostly depends on the $\vp$ and therefore on the viscosity.   
Although the dependence of multi-valued nature of velocities are very complicated but mostly depend on the viscosity, 
which we will see in the next figure}.
The solution corresponding to $r=4$ (solid red) is failed outflow (or fail to make transonic outflow solution and 
$\vr$ becomes again less than zero) due to very fast decreasing $\vp$ (panel \ref{fig4}c). 
Although $T$ is increasing but did  
not produce sufficient pressure gradient force to maintain $\vr$ positive and 
resulting matter falls back towards the BH. The solution corresponding to $r=4$ also feels more gravity than 
other solutions with higher values of $r$. 
{If we compare this kind of solution with no outflows analytical MHD solutions but $\vt=0$ for all $\theta$ \citep{zmay18}, 
in this case $\vp$ also decreases vary fast at high latitude. So $\vp$ plays a key role in generating the outflows. 
The black dashed curve behaves almost similar ways as the bipolar accretion outflow solutions are represented by \cite{xc97}.}

\begin{figure}[ht!]
\plotone{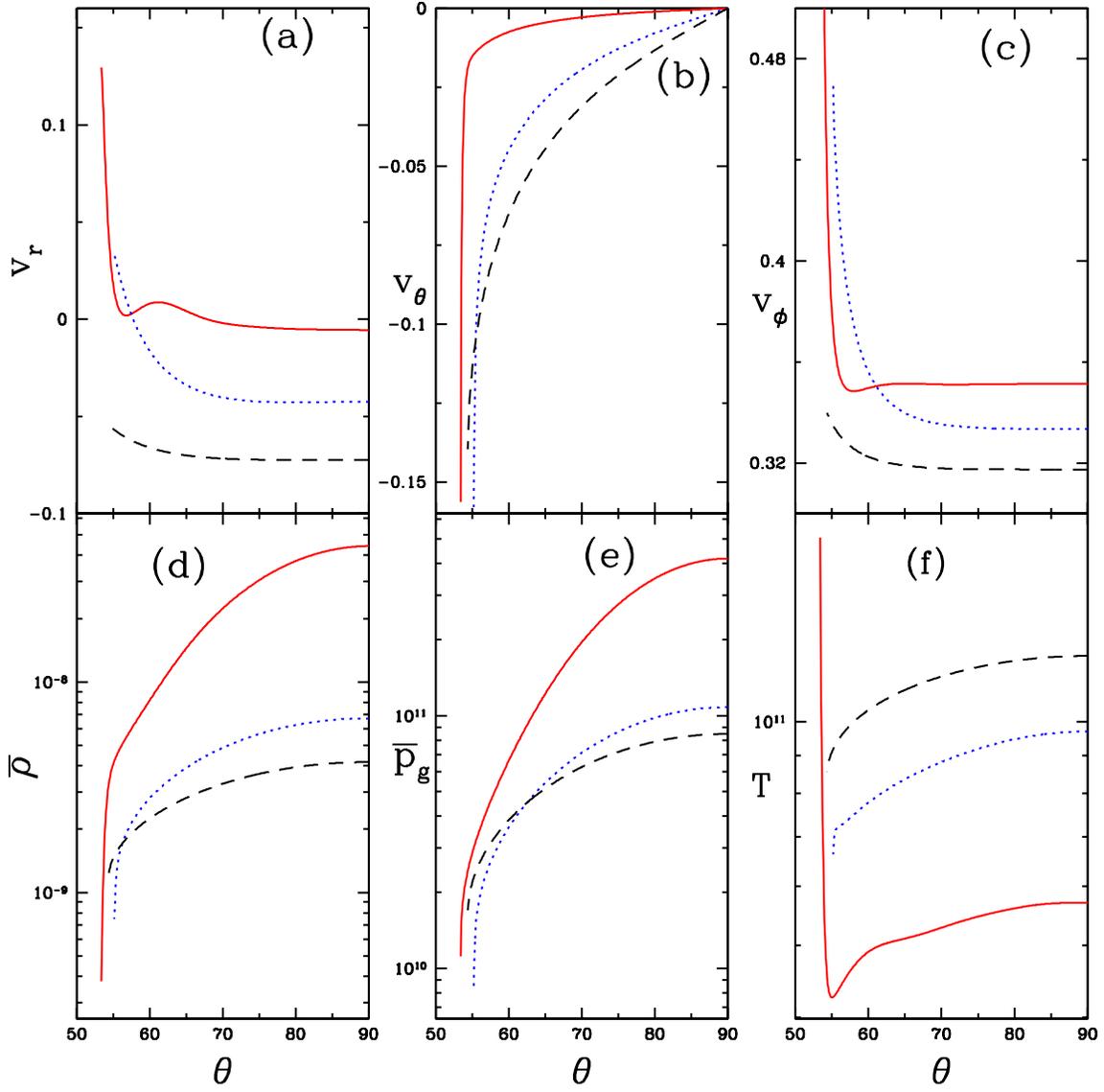}
\caption{Variations of flow variables with polar angle, $\theta$. Panels are showing variations of $\vr$ (a), $\vt$ (b), 
 $\vp$ (c), $\bar{\rho}$ (d), $\bar{p}_{\rm g}$ (e) and $T$ (f). 
These solutions are drawn for fixed radius, $r=6$ with different viscosity parameters, 
$\alpha_1=\alpha_2=0.01$ (solid, red), $0.1$ (dotted, blue) and $0.2$ (dashed, black) and
other parameters are corresponding from the second row of the Figure (\ref{fig3}).
\label{fig5}}
\end{figure}
The Figure (\ref{fig5}) is drawn for fixed radius $r=6$ 
with different viscosity parameters and used same parameters corresponding to the second row of the Figure (\ref{fig3}).
In panel (\ref{fig5}a), $|\vr|$ is high with high viscosity in the inflow region since $\lmda$ is low (Figure \ref{fig2}d) 
therefore $\vp$ is also low in panel (\ref{fig5}c). 
The $\vr>0$ is again high in the outflow region corresponding to same value of $\theta$ due to high acceleration for high viscosity, 
if we compare curves with solid red ($\alpha_1=\alpha_2=0.01$) and dotted blue ($\alpha_1=\alpha_2=0.1$). 
Since $|\vt|$ is also high for high viscosity (panel \ref{fig5}b), therefore the total outflow velocity ($v$) is high in both curves.
The $\vp$ (panel \ref{fig5}c) and $T$ (panel \ref{fig5}f) are monotonically increasing and decreasing, respectively for higher viscosity 
as compared with multi-valued curve for low viscosity (solid red). 
{Since higher viscosity makes flow more hotter as $T$ is higher (panel \ref{fig5}f) and also transports more angular momentum 
therefore, somehow which makes smooth variation of $\vp$ (panel \ref{fig5}c).} 
The solution corresponding to $\alpha_1=\alpha_2=0.2$ (dashed black) is not producing outflow due to low $\vp$,  
as all three solutions have same gravity pull at $r=6$.
The $\bar{\rho}$ (panel \ref{fig5}d) and $\bar{p}_{\rm g}$ (panel \ref{fig5}e) are decreasing smoothly with `$\theta$'. 
{Here $\bar{\rho}$ is higher for low viscous solution and therefore $\bar{p}_{\rm g}$ is also higher, 
since inflow $|\vr|$ is low.}
Interestingly, the gas density and pressure are not becoming zero at high latitude, specifically in high viscosity solutions because 
integrations are terminated at the outflow sonic surfaces.

\begin{figure}[ht!]
\plotone{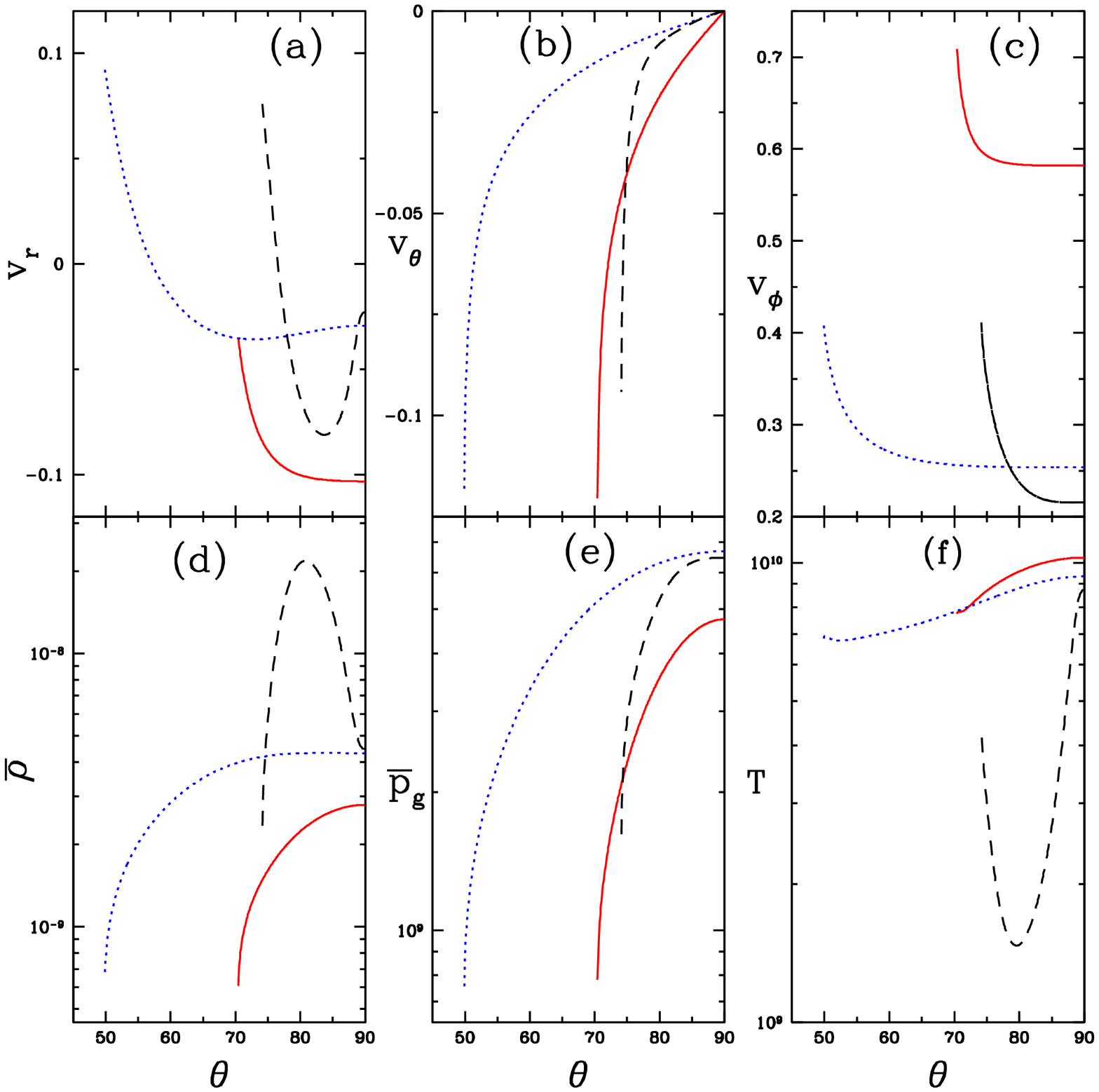}
\caption{Variations of flow variables with polar angle, $\theta$. Panels are showing variations of $\vr$ (a), $\vt$ (b), 
 $\vp$ (c), $\bar{\rho}$ (d), $\bar{p}_{\rm g}$ (e) and $T$ (f). 
These solutions are drawn with different radii, $r=3.1$ (solid, red), $8.5$ (dotted, blue) and $10.5$ (dashed, black) and
the disk parameters are same from the panel (c) of Figure (\ref{fig3}). 
\label{fig6}}
\end{figure}
Three curves of Figure (\ref{fig6}) are plotted from three different regions of the Figure (\ref{fig3}c), 
which are an inner (no outflow), middle (outflow) and outer (supersonic inflow $r>10\rs$) regions. 
The solution for no outflow (solid red, $r=3.1$) region is close to the BH and experience more gravity and 
the combined fluid centrifugal force and pressure gradient force 
are not sufficient to defend gravity. 
So, the matter is not able to cross the disk surface and always, $\vr<0$ (panel \ref{fig6}a).
The next middle with outflow region, solution (dotted blue, $r=8.5$) has appropriate forces, 
which make $\vr$ positive and give outflows. 
From outer part of the disk, the solutions from this region are showing very different behavior. 
If we look at dashed black curve (for $r=10.5$), as matter start expanding subsonically upward with increasing $|\vt|$ and 
$\vp$ along $\theta-$direction.  
Same time $\vr$ (panel \ref{fig6}a) and $T$ (panel \ref{fig6}f) are decreasing or increasing, simultaneously 
and resulting flow becomes supersonic and subsonic in the inflow region.    
In the same region $\bar{\rho}$ is also increasing or decreasing (panel \ref{fig6}d). 
In this region, velocities, density and gas pressure gradients are high from the solutions of other two regions 
and also cooler. 
This kind of inflow supersonic regions  
above the equatorial plane as seen in the second and third columns of the Figure (\ref{fig3}) except panel (f), 
which are surrounded by the black square symbol line. 
This kind of regions are not found with 
the low (panel \ref{fig3}d) or high (panel \ref{fig3}f) viscosity and low $r_{\rm t}$ (panel \ref{fig3}g) solutions.
Here the disk viscosity parameters roughly categorized as $\sim 0.01$ is low and $\gtrsim 0.1$ is high. 

{The outflow solutions have a good qualitative agreement with radial self-similar adopted 2D HD \citep{XW05,JW11} and 
MHD \citep{sa16,mby16} flows. 
Since the disk vertical thickness of our solutions depends on the radius and flow parameters. 
So there is a possibility that the disk thickness can matched with previous studies for some suitable flow parameters. 
If we compare results for 2D disk structure with no outflow \citep{ny95a,zmay18} then 
the our disk vertical thickness is low since the integration is terminated at the sonic surface. 
For no outflow disk structure (Figure \ref{fig3}f), the thickness increased with the increasing viscosity. 
Moreover, the nature of flow variables profiles along the $\theta-$ direction are mostly consistent 
with simulation by \cite{yyob14}.}
\subsubsection{effect of $\ttp$ and $s$} \label{subsubsec:efttps}
\begin{figure*}
 \gridline{\fig{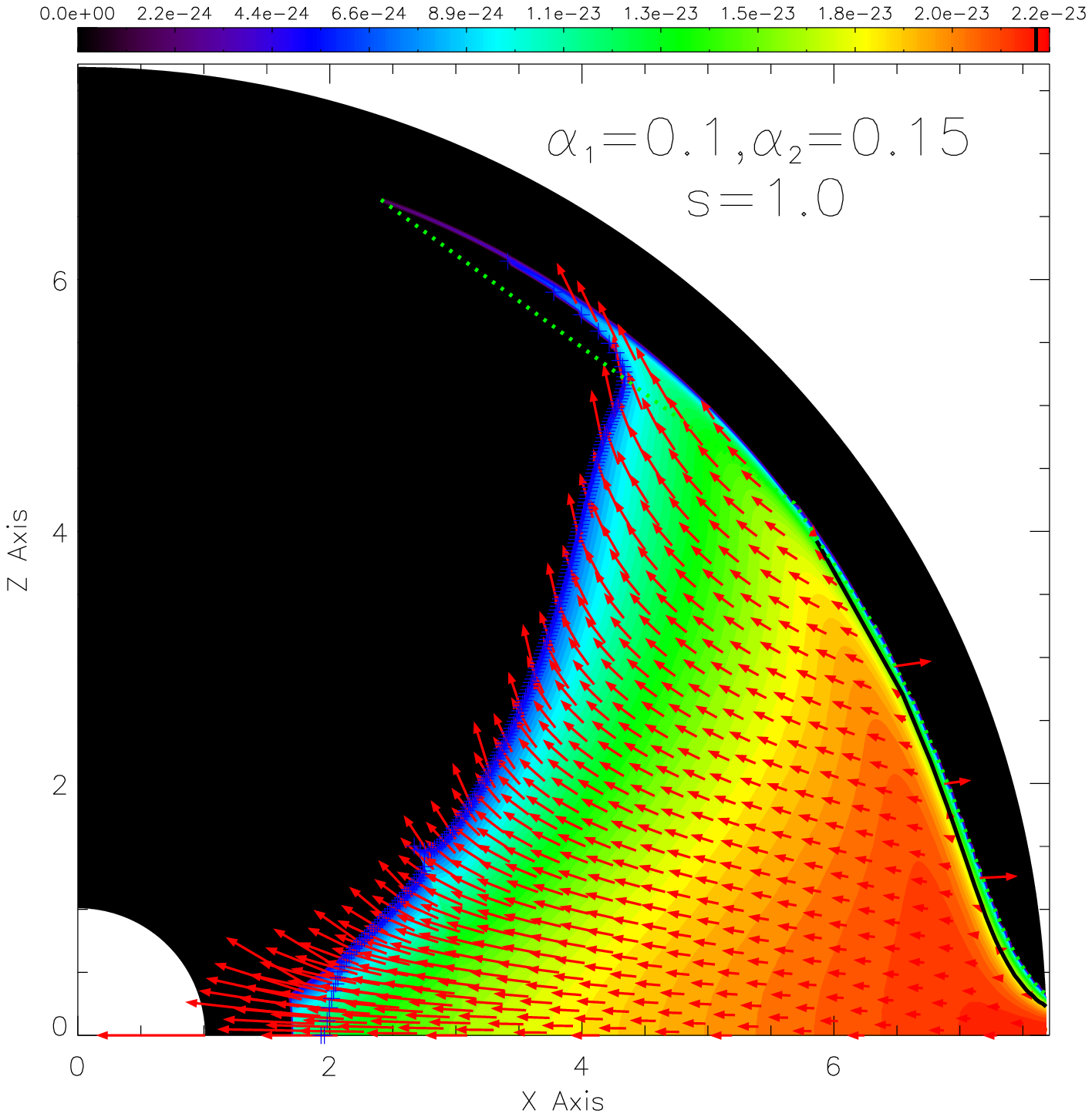}{0.3\textwidth}{(a)}
          \fig{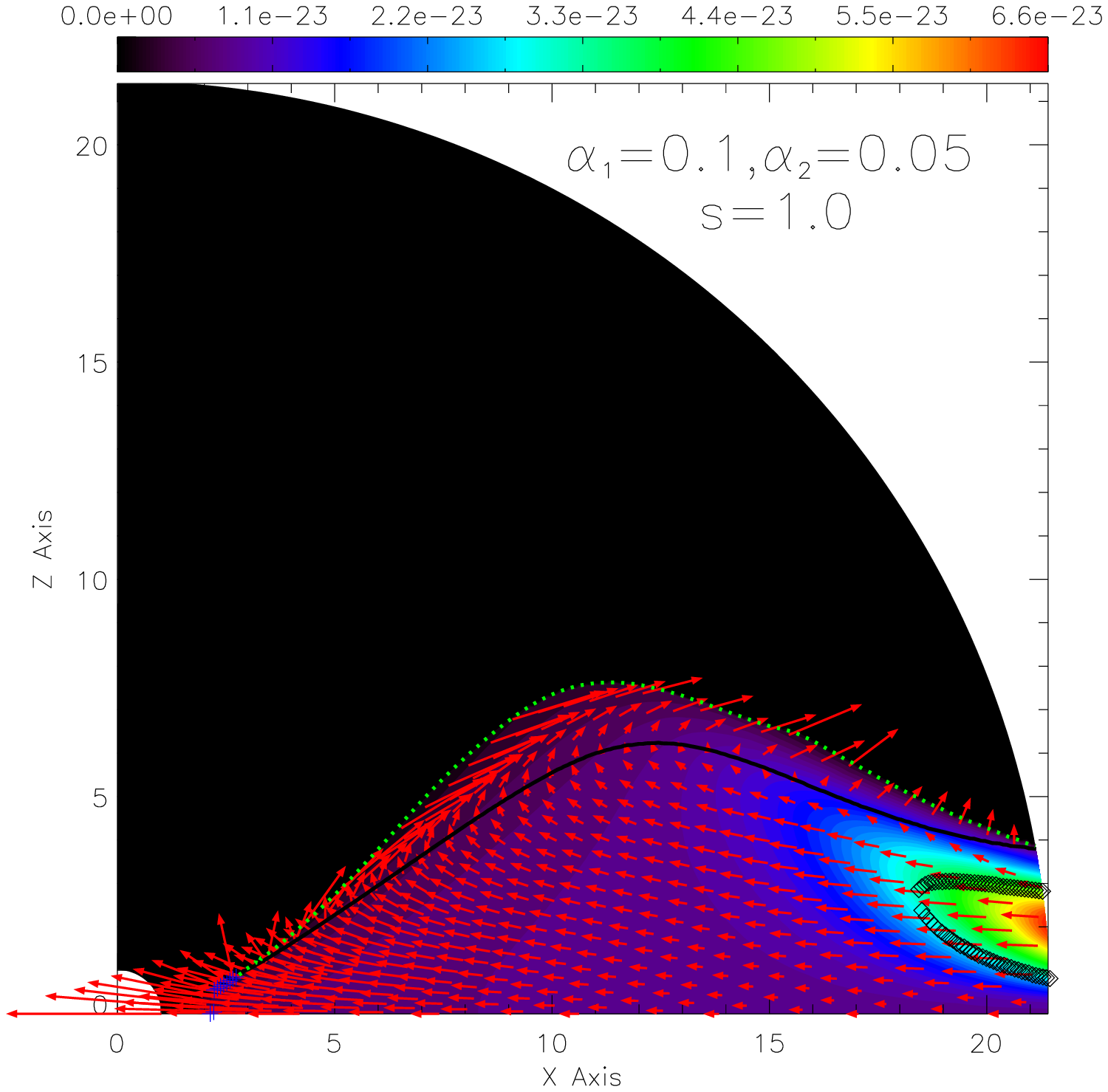}{0.3\textwidth}{(b)} 
          \fig{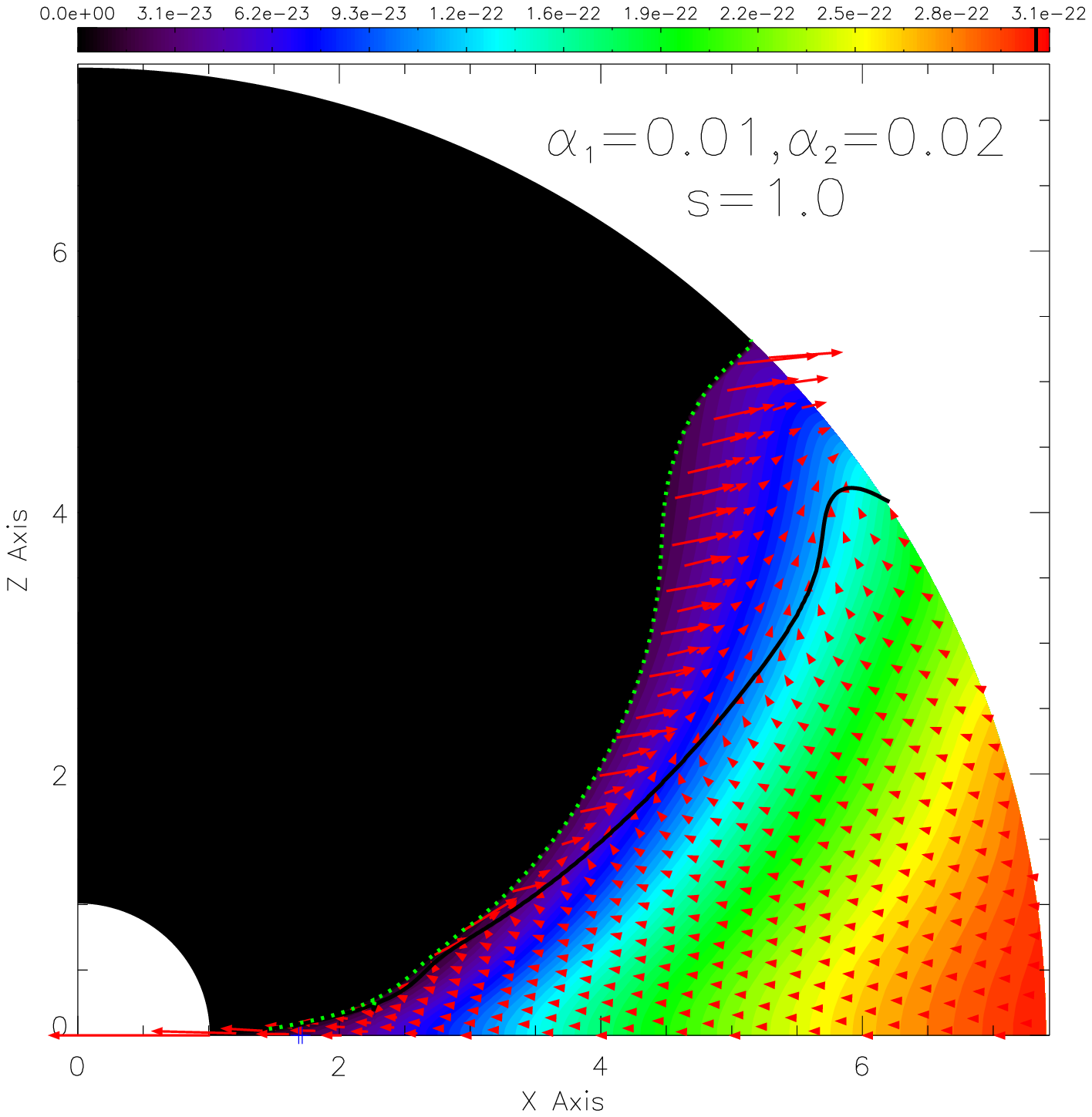}{0.3\textwidth}{(c)}      
 }
 \gridline{\fig{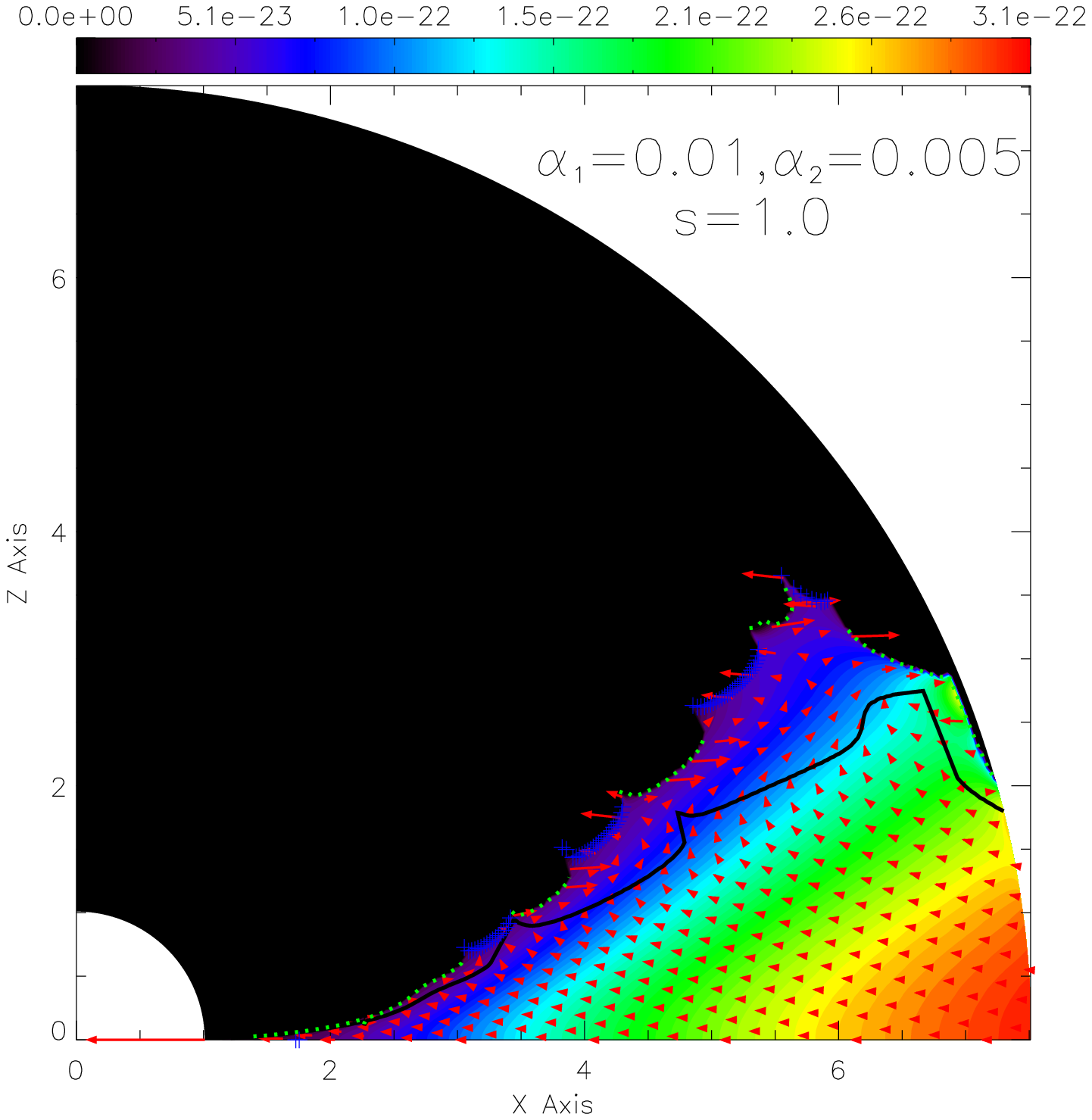}{0.3\textwidth}{(d)}
           \fig{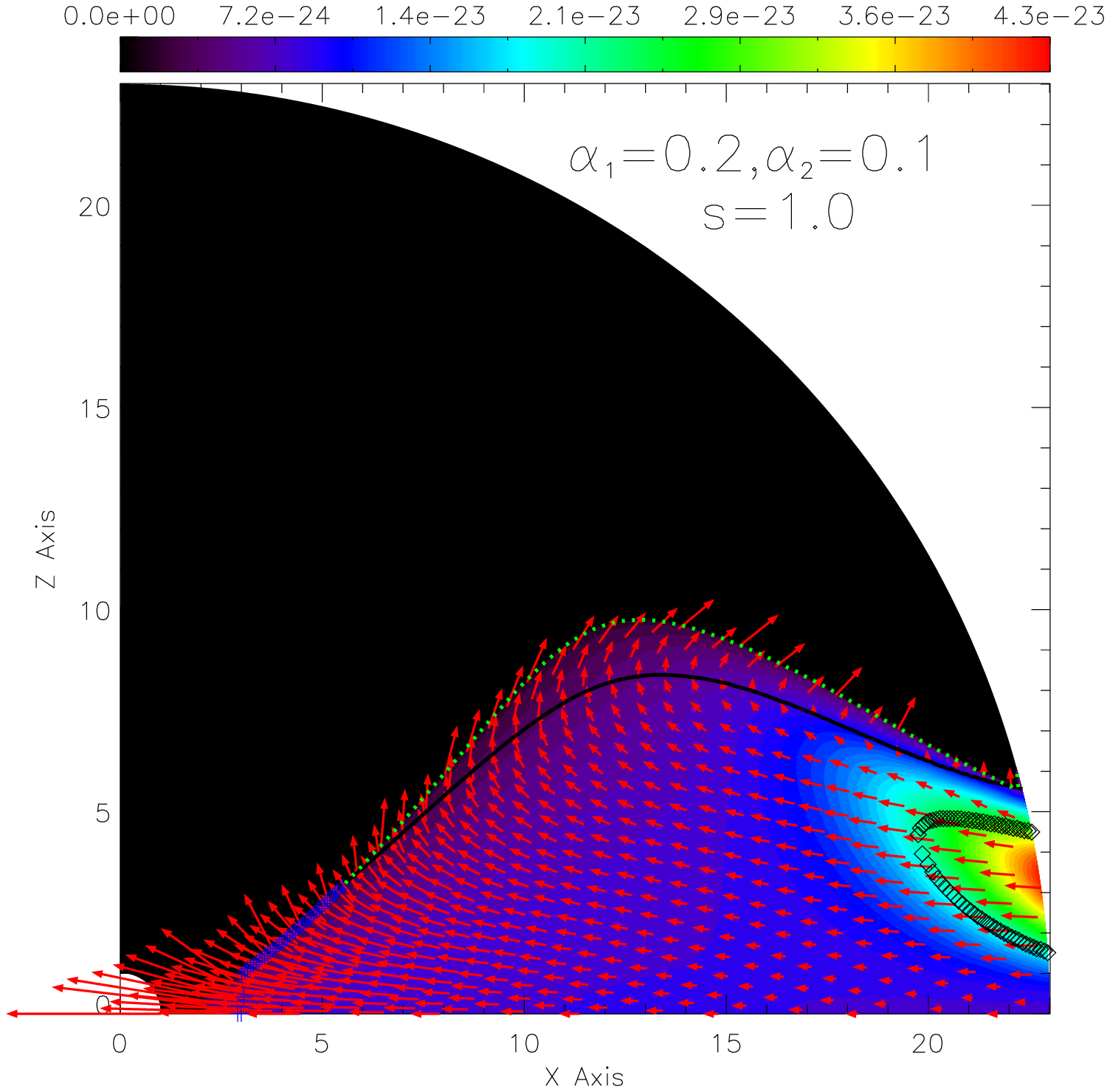}{0.3\textwidth}{(e)}
          \fig{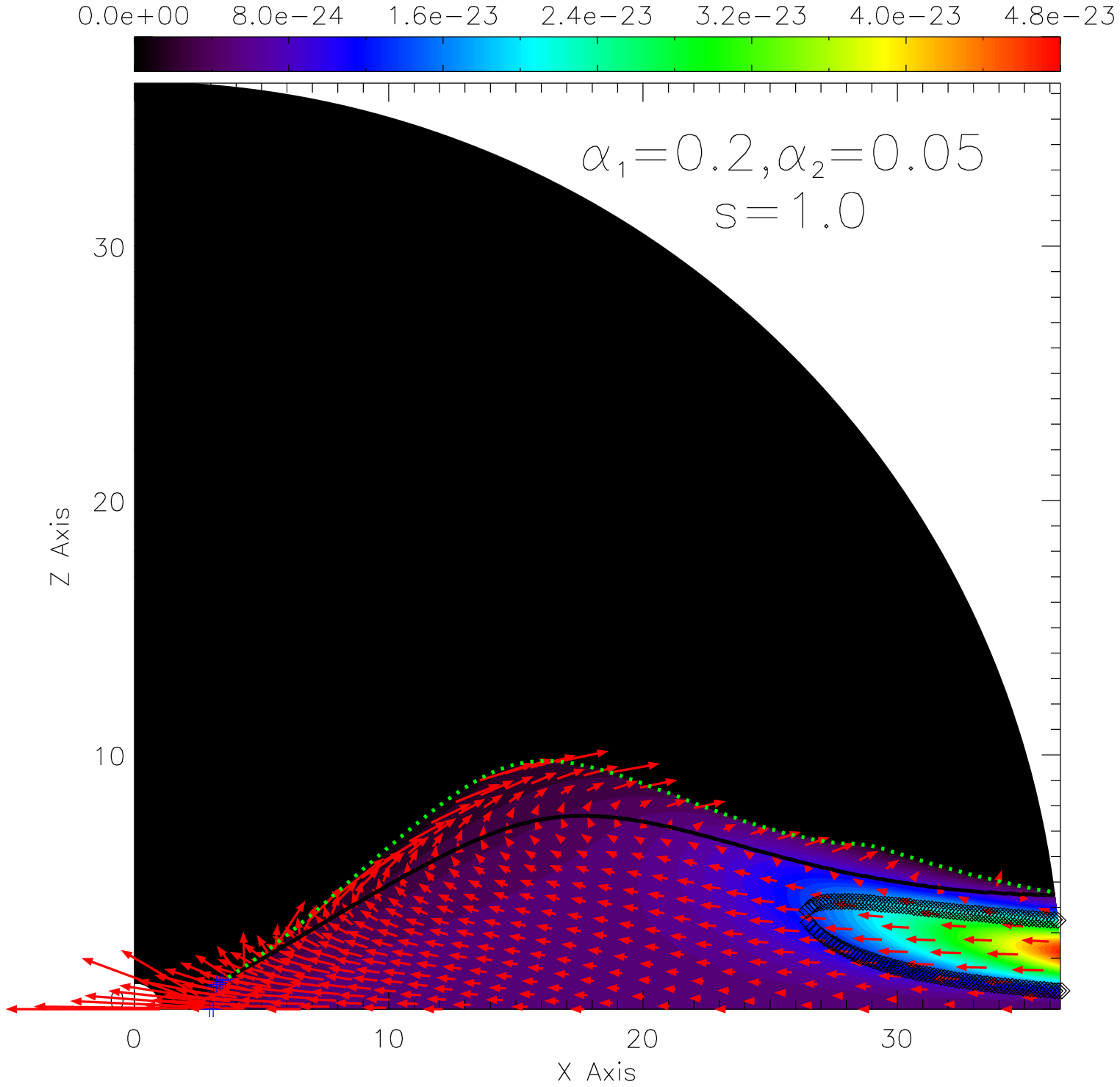}{0.3\textwidth}{(f)}
 }
 \caption{The density contours and velocity fields are plotted with same disk parameters corresponding to the first panel, 
 panel (d) and panel (f) of the Figure (\ref{fig3}).
 Here figures have plotted with different values of $\alpha_2$ as mentioned in each panel.
 }
 \label{fig7}
\end{figure*}

Now we are studying in this subsection effects on the 2D disk structure with variation of $\alpha_2$ 
for $\ttp$ and $s$.
Here we are taking three cases, one from the Figure (\ref{fig3}a) which is plotted with $\gamma=\geff=5/3$, 
second, which has low viscosity with weak outflows from the Figure (\ref{fig3}d) and 
third, high viscosity with no outflows from the Figure (\ref{fig3}f). 
All the cases are presented in Figure (\ref{fig7}) with variation of $\alpha_2$ and 
keeping other parameters fixed for each cases.
The first case of the Figure (\ref{fig3}a) is represented with two viscosity parameters, 
$\alpha_2=0.15$ (panel \ref{fig7}a) and $0.05$ (panel \ref{fig7}b). 
The panel (\ref{fig7}a) has disk structure with no outflows because more transfer of angular momentum due to high $\alpha_2$ and 
as a result $\vp$ decreased much at high latitude and flow variables variation became similar to the dashed black curve of the 
Figure (\ref{fig5}). So, all the matter will fall supersonically onto the BH after crossing the sonic surface. 
In the panel (\ref{fig7}b) is plotted with $\alpha_2=0.05$, which gave outflow solutions and 
outflow region also increased from the Figure (\ref{fig3}a).
The second case from the Figure (\ref{fig3}d) is represented with $\alpha_2=0.02$ (panel c) and 
$\alpha_2=0.005$ (panel d) of the Figure (\ref{fig7}). 
When we compared with the Figure (\ref{fig3}d), the outflow region and strength are increased in panel (\ref{fig7}c) but decreased in panel (\ref{fig7}d).  
In the panel (\ref{fig7}c), $\vt$ is very small from the value of $\vr$, 
so velocity vectors are almost parallel to the equator and the matter seems going back at high latitude.
The last case of the Figure (\ref{fig3}f) is again drawn with two different viscosities, $\alpha_2=0.1$ (panel \ref{fig7}e) and $0.05$ (panel \ref{fig7}f).
In both panels \ref{fig7}(e and f), outflow region is increased with decreasing $\alpha_2$ from the Figure (\ref{fig3}f).
From this study, we can say that disk structure depends on $\ttp$ and also depends on value of viscosity parameters. 
Since for low viscosity ($\alpha_1$), say $\sim 0.01$, the outflows are increasing with increasing $\alpha_2$ and 
for high viscosity ($\alpha_1$), say $\gtrsim 0.1$, 
the outflows are increasing with decreasing $\alpha_2$, when keeping $\alpha_1$ fixed. 

\begin{figure}
\plotone{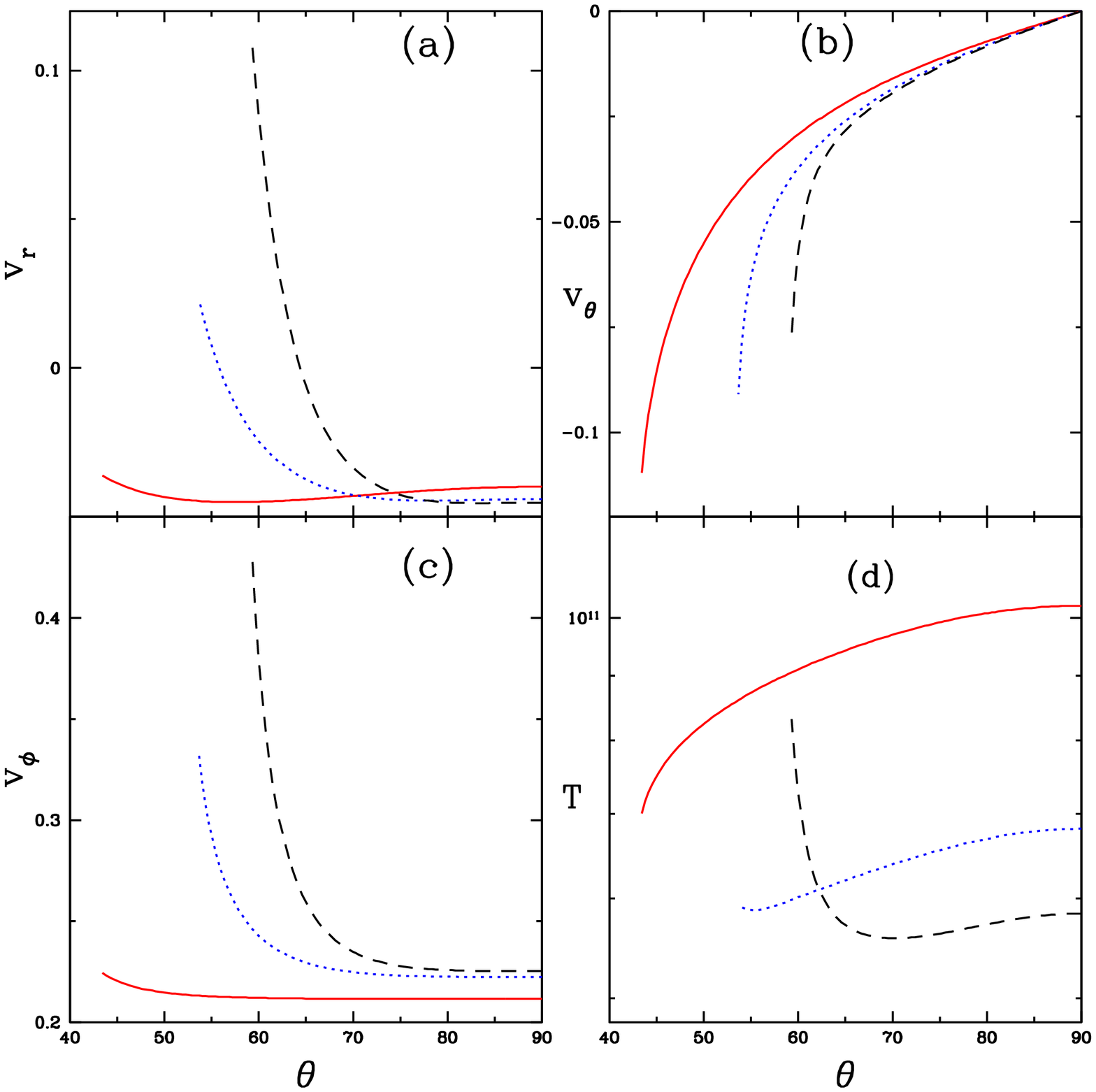}
\caption{These solutions are plotted with changing viscosity parameter, $\alpha_2=0.2$ (solid red), $0.1$ (dotted blue) 
and $0.05$ (dashed black) at same $r=10$ and other parameters are same as in the panel (f) of Figure (\ref{fig3}).
\label{fig10}}
\end{figure}
In Figure (\ref{fig10}), we are represented variation of three velocities and flow temperature with $\theta$ 
at fix $r=10$ for different $\alpha_2$ values, which are taken from the Figures (\ref{fig3}f, \ref{fig7}e and \ref{fig7}f). 
For $\alpha_2=0.2$ (solid red) is not showing the outflow. Since 
variation and values of $\vp$ is less (panel \ref{fig10}c) and $T$ is also decreasing above the equatorial plane (panel \ref{fig10}d), 
therefore combine effect of the outflows driving forces are not enough to make $\vr$ positive. 
When we decreased $\alpha_2=0.1$ (dotted blue) and $0.05$ (dashed black) then the $\vp$ is high and rising faster at high latitude (panel \ref{fig10}c). 
So $\vr$ becomes positive and gives outflow. 
The similar behavior of $\vp$ has also found in the simulation with the inclusion of $\ttp$ \citep{yyob14}, 
which decreases $\vp$ at high latitude.
Moreover, $\vr$ (panel \ref{fig10}a) and $\vt$ (panel \ref{fig10}b) are increasing faster with decreasing $\alpha_2$, 
so the outflow strength is also increased. 

\begin{figure}
\plottwo{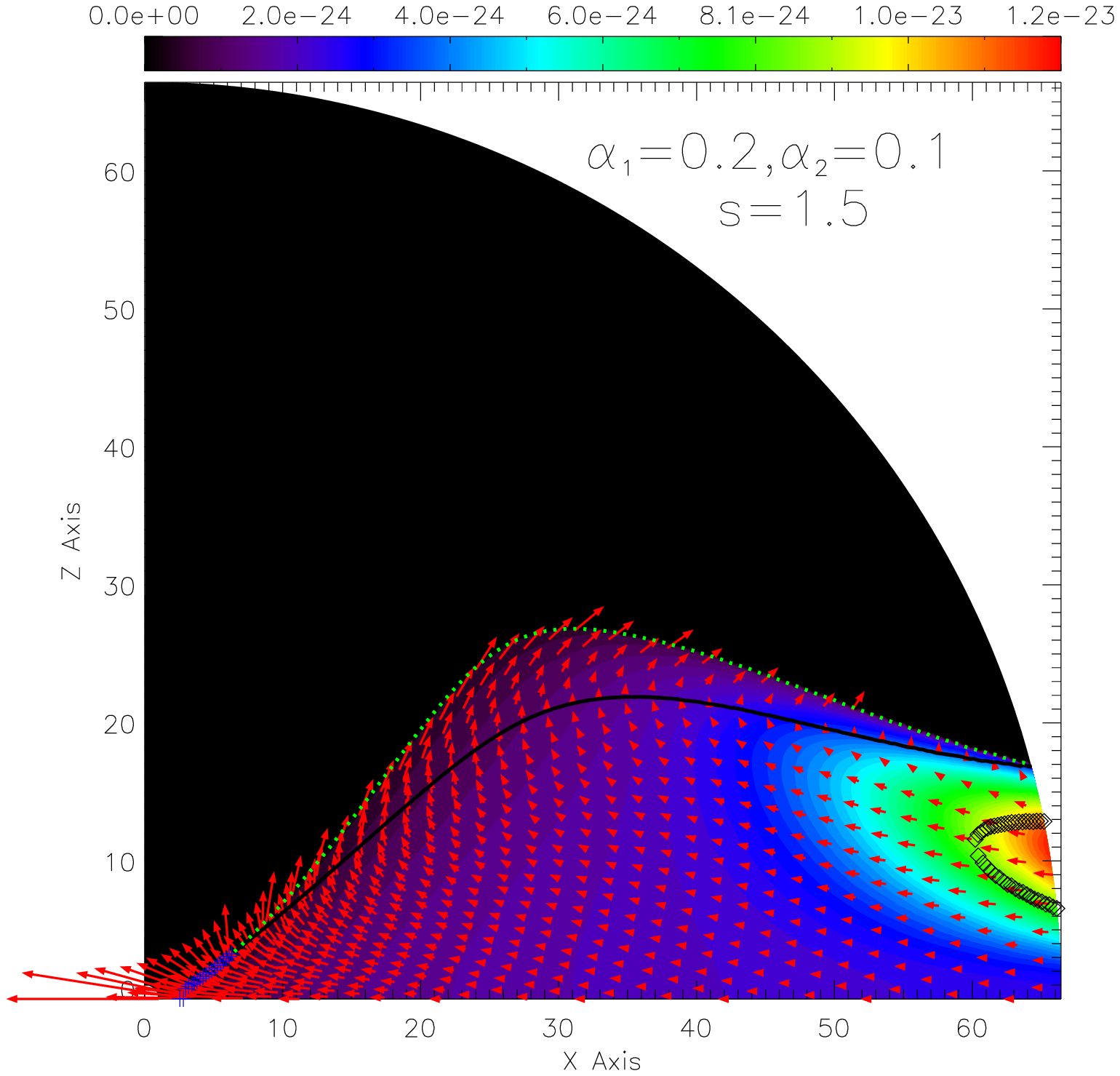}{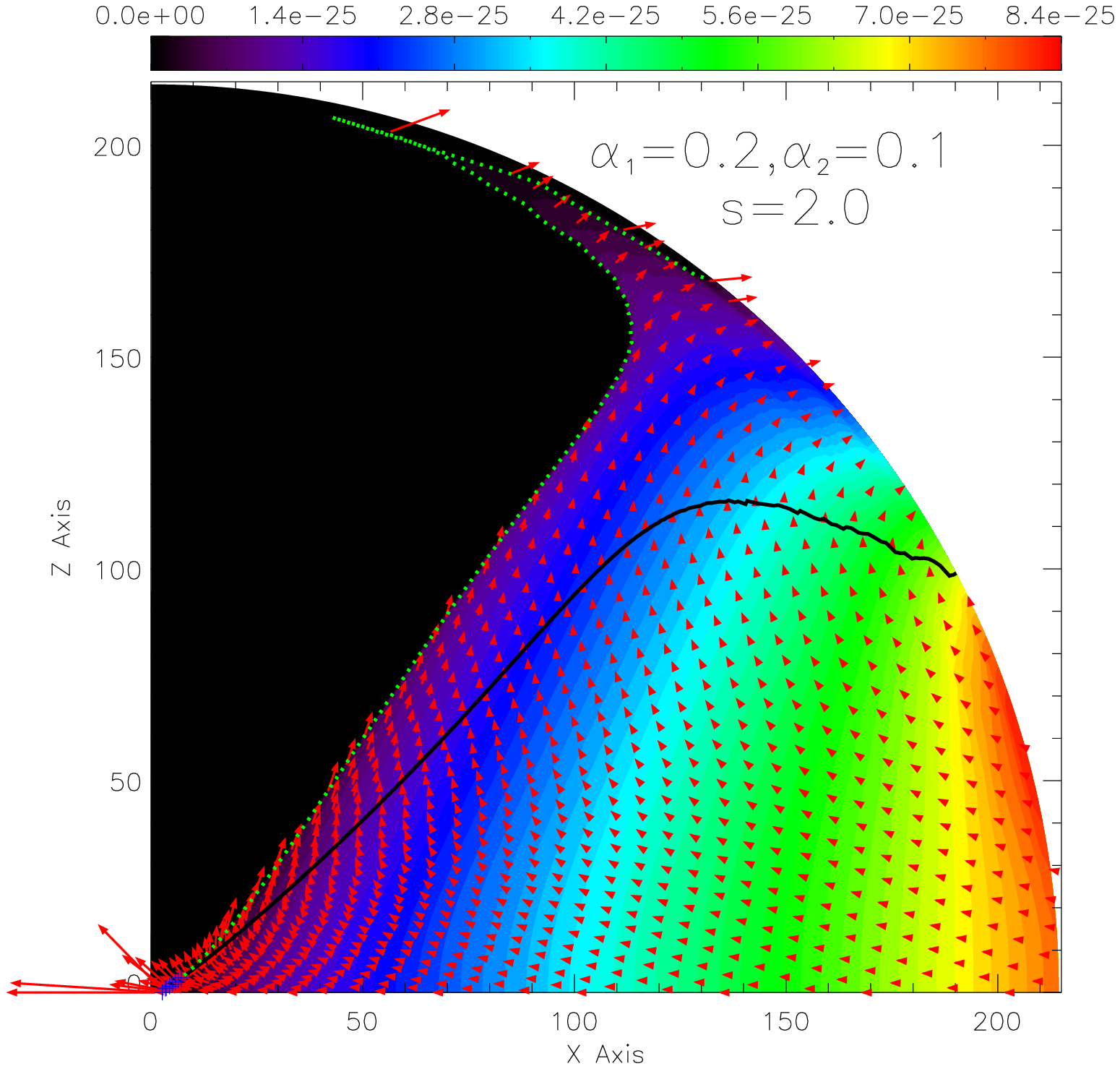}
\caption{The density contours and velocity fields are presented with different $s$ and 
other disk parameters are same as in the panel (e) of Figure (\ref{fig7}).
Here value of $s$ is mentioned in both panels with viscosity parameters.
\label{fig8}}
\end{figure}

Now, we are changed value of $s$ with keeping other parameters fixed and studied effects on the disk structure. 
Here we are taken case of the Figure (\ref{fig7}e) with changing $s$ and presented in Figure (\ref{fig8}). 
In both panels of the Figure (\ref{fig8}), the outflow region and strength are increased with increasing $s$. 
Since the outflows are much affected by the viscosity parameters and mass-loss parameter. 
Therefore, we want to see the variation of local energy of the inflow-outflow and definition of the local energy of the flow is 
\begin{equation}
 B(r,\theta)=B=\frac{\vr^2}{2}+\frac{\vp^2}{2}+\frac{\vt^2}{2}+h+\Phi.
 \label{bth.eq}
\end{equation}
This is modified Bernoulli energy parameter for the 2D flow. 
Which is similar to the local energy defined in the appendix (\ref{sec:Feqneq}) as 
the Bernoulli parameter $\be$ on the equatorial plane, when $s$ is zero. 
Here, $h$ is specific enthalpy. 

\begin{figure}
\plotone{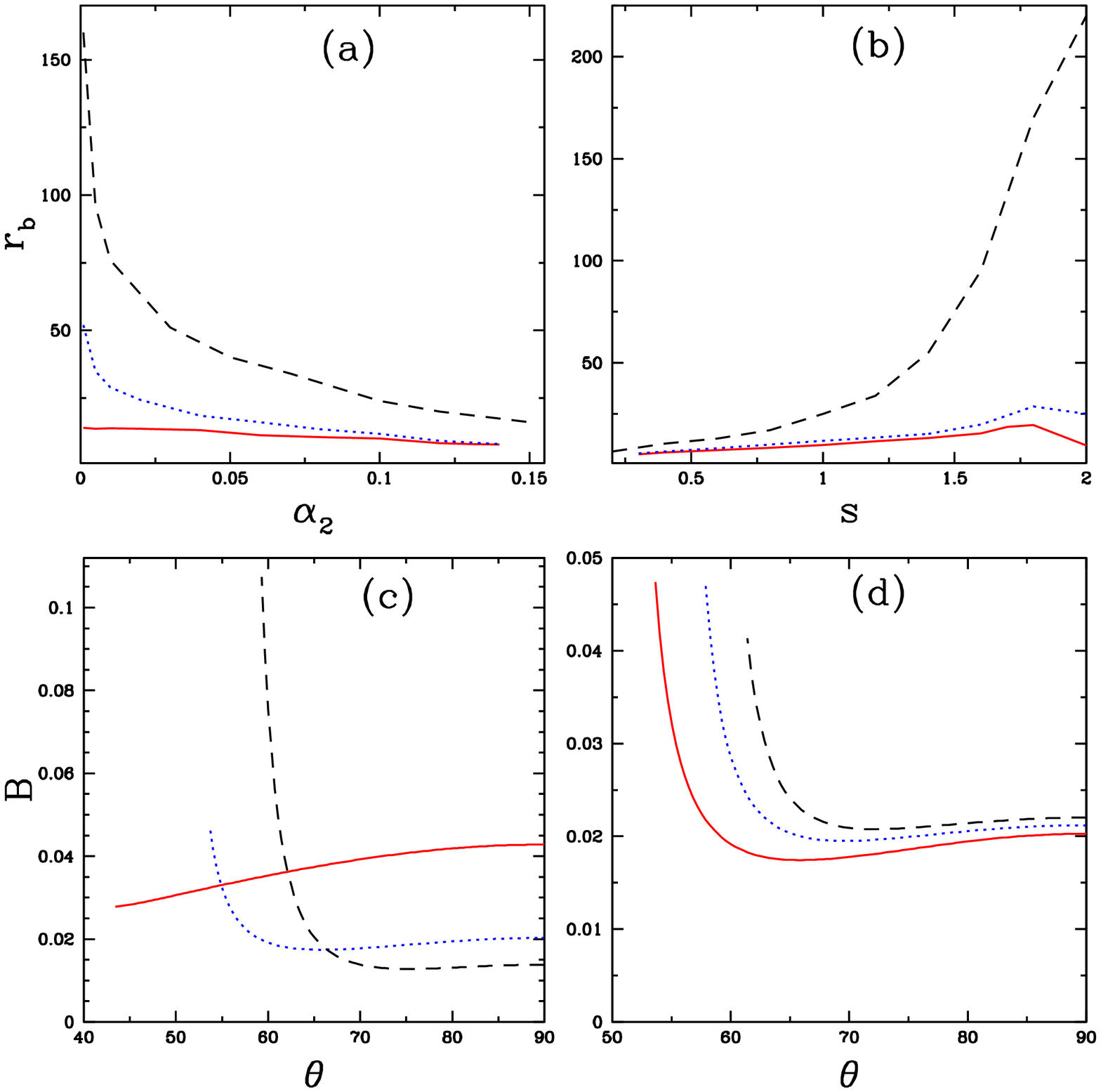}
\caption{Variation of $r_{\rm b}$ with $\alpha_2$ (panel a) and $s$ (panel b) are presented. 
In the panel (a), curves are presented with fixed $s=1$ and other disk parameters are 
${E}=-0.02, \alpha_1=0.1$ (solid red), ${E}=-0.001, \alpha_1=0.1$ (dotted blue) and ${E}=-0.001, \alpha_1=0.2$ (dashed black). 
In the panel (b), curves are presented with fixed $\alpha_2=0.1$ and other disk parameters are same as the panel (a). 
Panels (c) and (d) are presented variations of $B$ with $\theta$ at fixed $r=10$ for parameters ${E}=-0.001, \alpha_1=0.2$. 
The panel (c) is plotted with different, $\alpha_2=0.2$ (solid red), $0.1$ (dotted blue) and $0.05$ (dashed black) for fixed $s=1$.
The panel (d) is plotted with different, $s=1.0$ (solid red), $1.5$ (dotted blue) and $2.0$ (dashed black) for fixed $\alpha_2=0.1$.
\label{fig9}}
\end{figure}
In Figure (\ref{fig9}), we are presented the variations of outer boundary of the outflows ($r_{b}$) with $\alpha_2$ (panel \ref{fig9}a), $s$ (panel \ref{fig9}b) and 
Bernoulli parameter $B$ with $\theta$ in panels \ref{fig9}(c and d) and other details are written in the caption. 
The two curves, solid line (red) and dotted line (blue) in the panel (\ref{fig9}a) are represented with different flow 
constant of motion ${E}=-0.02$ and $-0.001$, respectively. 
So, they are have different $r_{\rm t}=28$ for ${E}=-0.02$ and $705$ for $E=-0.001$. 
The curve with lower $r_{\rm t}$ is has small outflow region and which 
more clear towards lower values of $\alpha_2$ in the panel (\ref{fig9}a). 
Here $r_{\rm b}$ is higher means more matter going out from the disk or higher mass outflow rate. 
Again in the same panel (\ref{fig9}a), another curve with dashed (black) line is plotted with the same ${E}$ as the curve dotted (blue) but both have 
different $\alpha_1=0.1$ and $0.2$. Other flow parameters are same for both the curves. 
We found that the outflows are high with higher $\alpha_1$ for same $\alpha_2$. 
Since high $\alpha_1$ rises more temperature and kinetic energy, therefore local specific energy of the flow is increased as seen in the Figure (\ref{fig2}f). 
Here outflow region is increased with decreasing $\alpha_2$ but disk thickness is decreased as seen in the Figure (\ref{fig7}). 
For high $\alpha_2>0.15$, we did not find the outflows but has the inflow 2D disk structure as seen in the Figure (\ref{fig3}f).  
In the panel (\ref{fig9}b), the outflow region is increased with increasing $s$ and $\alpha_1$. 
The two curves, solid and dashed line are become maximum around $s\approx1.8$ and decreased with further increasing $s$. 
Since for $s\gtrsim 1.8$, the gas pressure variation along the radial direction becomes almost flat on the equatorial plane. 
In panels (\ref{fig9}c), curves with variations of the $B$ are plotted for same solutions of the Figure (\ref{fig10}). 
The solid red curve ($\alpha_2=0.2$) is decreased with decreasing $\theta$ and gave no outflow solution as in the (\ref{fig3}f). 
Other two curves ($\alpha_2=0.1$ and $0.05$), the $B$ are increased with decreasing $\theta$ at high latitude and gave outflows. 
In panel (\ref{fig9}d), the $B$ is increased with decreasing $\theta$  and increasing $s=1.0$ (solid red), 
$1.5$ (dotted blue), $2.0$ (dashed black). So, the outflow region and strength are increased with increasing $s$ 
as seen in the Figure (\ref{fig8}). 
\section{SUMMARY AND DISCUSSION}
We have explicitly obtained radial fluid equations (\ref{dt1.eq}-\ref{dl.eq}) on the equatorial plane and 
used them with symmetric conditions (\ref{bc.eq}) for solving ODEs (\ref{rc2.eq}-\ref{eg2.eq} and \ref{ce3.eq}) along the $\theta-$ direction. 
First, we obtained radial flow variables with its derivatives by integrating the radial fluid equations. 
Second, we integrated ODEs along polar direction by using obtained polar flow variables at $\theta=\pi/2$. 
These two integration are run one by one
at each step of $r$,  
after repeatedly doing so, we got complete 2D disk structure of the flow.   
We found two distinct regions in the 2D flow for the viscosity $\alpha_1>0.01$, 
one is the inflow region when $\vr<0$ around the equatorial plane and second is the outflow region when $\vr>0$ 
above the inflow region. Both regions are separated by the disk surface with $\vr=0$. 
For $r\lesssim 4$, we found only the inflow region and flow is supersonic. 
For low viscosity $\alpha_1\lesssim 0.01$, we also found failed outflow regions in the outflows part 
above the disk surface. The failed outflows means the 
flow radial velocity again becomes negative ($\vr<0$) at high latitude. 
Here the outflow regions are plotted up to the sonic surface (when $M=1$). 
Since integration is encountered the problem due to discontinuity at $M=1$ in solving the equations along the polar direction. 
The inflow disks are also showing supersonic regions just  
above the equatorial plane and it appears away from the BHs $r>10$, 
which depends on the flow parameters (the second and third columns of Figure {\ref{fig3}} except panel f, and Figure {\ref{fig7}b, e \& f}). 
This region is surrounded by sonic surface in the disk but does not has proper critical points or 
discontinuity in the flow means integration passes smoothly at this surface. 
This kind of supersonic regions in the inflow part are not formed, 
when the disk having no outflows with high $\alpha_2>0.15$ (Figures \ref{fig3}f, \ref{fig7}a) or 
 low $\alpha_1\lesssim 0.01$ (Figures \ref{fig3}d, \ref{fig7}c \& d) or 
$s>1.7$ (Figure \ref{fig8}), when the disk having outflows.

Our results are having the inflow and outflow regions for a certain range of the viscosity parameters 
and they also depend on other disk parameters. 
Which is consistent with some analytical \citep{xc97,XW05,JW11,g12,g15} and 
numerical simulation studies \citep{omnm05,otm07,ywb12,ybw12,yyob14,Jetl15,ygnsbb15}. 
Although, our solutions and the disk structure are quite different from 
the previous analytical studies on 2D disk flow with self-similar conditions,  
\eg the size and shape of the disks, behavior of the solutions and supersonic regions. 
But basic properties and some of the solutions are similar qualitatively. 
This is clear due to differences in boundary conditions on the equatorial plane 
and used two viscous stress components in our model fluid equations of motion.
In self-similarity, all the radial flow variables (velocities and sound speed) are same so the Mach number is one and 
constant at every radius but in our case, this happened only at the critical point of the transonic ADAF solutions 
(Figure \ref{fig2}). 
Moreover, values of these radial variables have not changed with viscosity parameter and $\beta$ as 
in paper \citep{JW11}, which is unlike with our case. 
Due to above reasons, we got outflow structure is different from other analytical 2D flow studies. 
For most of the common flow parameters of my studies are showing outflows, \eg range of the $s$ from $0.2$ to $2$,  
range of $\alpha_1$ from $0.01$ to $0.2$ and $\alpha_2 \lesssim 0.15$. 
Although the lower range of the $s$ depends on the $\alpha_1$ but upper limit $s=2$ is used here. 
Since, the gas pressure profile becomes flat along radial direction on the equatorial plane for $s\sim 1.8$. 
For high $\alpha_2>0.15$, we did not get outflow solutions but having 2D advective disk structure (Figure \ref{fig3}f 
\& \ref{fig7}a). 
Moreover, for high $\alpha_1>0.1$ and low $\alpha_2<0.01$ with high $s>1$, 
we may get larger outflow region or almost from whole ADAF disk, which can be predicted from study of the Figure \ref{fig9}(a \& b). 
The outflow region is also larger for high ${E}$ of the flows 
(the last row of the Figure \ref{fig3} and \ref{fig9}a \& b). 
The disk structure is not much affected by the variations of $\gamma$ or $\beta$ but 
the outflow strength is high for radiation-dominated flows ($\beta<1$). 
{Although the main feature of the outflows are roughly consistent with the simulation results but 
the vertical thickness, the inflow and outflow regions of our most of the solutions are small from the simulations by 
\cite{ywb12,ybw12,yyob14} with using two non-zero azimuthal components of the anomalous shear stress tensor. 
These differences may arise due to analytical approach with assuming constant mass-loss parameter ($s$) throughout the disk, 
the integration problem after the sonic surfaces and assumed boundary conditions at the equatorial plane. 
Moreover, the radial power law index of $a_{\rm s}$ and $\vp$ around the equatorial plane 
are roughly close to the simulation \citep{ywb12}. 
Although, the scaling for $\vr$ and $\rho\propto r^{s-2}/\vr$ are mostly depend on the $r_{\rm t}$ (or $E$) as explained in the 
Figure (\ref{fig2}) and $s$, respectively.
}

The outflows are more favorable for $\alpha_2\ge\alpha_1$ when low $\alpha_1\sim 0.01$ and 
$\alpha_2<\alpha_1$ when high $\alpha_1\gtrsim 0.1$ as cases have presented here.  
All disk parameters may have outflow solutions but need to find suitable outflow structure parameters, like, $\alpha_2$ and $s$.
Similarly, we also found that for $\alpha_2>0.15$, which has the 2D disk structure 
but no outflows for any value of $s$. 
Since $s$ and $\alpha$ are fixed here for a particular solution, therefore we have varied possible values of $s$ and 
$\alpha$ for the 2D disk structure. 
When doing this we found range of outflows region for $s$ and $\alpha_2$ (Figure \ref{fig9}).
We have 2D disk structure and no outflows for $s\neq 0$ means no mass loss from the disk or 
matter is bound with the disk even $\vt$ is non-zero with $\vr<0$. 
The transonic surface formed for the outflows ($\vr>0$, above the inflow disk surface), 
at this surface outflow density $\gtrsim 20\%$ from the local equatorial plane density. 
So the outflow solutions can cross the transonic surface and give supersonic outflows but 
the present analytical study is limited to the outflow sonic surface.  

The present study is done with one kind of input accretion solutions (ADAF-thin) 
on the equatorial plane for calculation of the 2D disk structure. 
We found that only inner region of the advective disk is participating in the outflow 
generation and size is around a few tens of the Schwarzschild radius.  
Which is consistent with the observed size of the outflow region around M87 \citep{jbl99,detal12}. 
Our studies are  
also supporting the two zone configuration theory of the fluid flows \citep{emn97,ds13}, since outflow region and strength are changed with 
changing viscosity parameters, $r_{\rm t}$ or ${E}$ and $s$. 
The $\ttp$ component of viscosity is affecting the outflows and decreasing angular momentum above the equatorial plane (Figure \ref{fig10}), 
which has also seen in the simulation \citep{yyob14} and variations of the polar flow variables are also roughly consistent with 
this simulation. 
This kind of analytical studies is worth pursuing for detail studies of flow solutions and the 
disk structure with various flow parameters that 
characterized the flow. This study also gives the idea about no outflows (for very low or high viscosity), 
outflows (for moderate viscosity) and 
failed outflows (for low viscosity). 
Incidentally, the shape of our disk structure are similar to the variations of $\be$ along the radial direction (last row of Figure \ref{fig2}), 
which is supporting the idea of positive local energy for generation of the outflows \citep{ny95a,BB04}. 
Moreover, we also found the supersonic region in the inflow away from the BH before the outflows happening in some cases.
At the boundary of this region, as matter moved inward with sharply rising temperature and decreasing   
bulk velocity along the radial direction  
with resulting flow makes a transition from supersonic to subsonic. 
This kind of transitions 
may have possibility of shocks in the inflows as 
studied by many authors \citep{c89,bdl08,lckhr16,kc17}.

This kind of studies and techniques for solving ODEs can be worth pursuing
in future. One can use variety of advective solutions on the equatorial plane, such as an ADAF-thick disk \citep{lgy99}, 
slim disk \citep{acls88} and shock solution \citep{kc14,kc17} 
with relevant cooling mechanisms in the study of two-dimension flows and also compare them.  
\acknowledgments
We thank Tuan Yi for helpful discussion. This work was supported by the National Basic Research Program of China
(973 Program) under grant 2014CB845800, and the National Natural Science Foundation of China under grants 11573023 and 11333004.
We also thank the anonymous referee for their helpful comments and suggestions. 
%




\appendix

\section{Fluid equations on the equatorial plane}
\label{sec:Feqneq}
Here we made two more assumptions in order to solve fluid equations on the equatorial plane, one 
all ${\partial}/{\partial\theta}=0$ and other
$\vr=\vre, \vt=0, \vp=\vpe, \Theta=\Thre, \rho=\rhe$  
then equations (\ref{ce.eq}-\ref{eg.eq}) written as,
the continuity equation,
\begin{equation}
 \frac{1}{r^2}\frac{d}{d r}(r^2\rhe \vre)=0
 \label{ce1.eq}
\end{equation}
Navier-Stokes equations are $r-$ component,
\begin{equation}
 \vre \frac{d \vre}{d r}-\frac{\vpe^2}{r}+\frac{1}{\rhe}\frac{d \Pe}{d r}-F_r=0
 \label{rc1.eq}
\end{equation}
$\phi -$ component
\begin{equation}
 \vre \frac{d \vpe}{d r}+\frac{\vpe}{r}\vre=
 \frac{1}{\rhe r}[\frac{1}{r^2}\frac{d}{d r}(r^3\trpe)]
 \label{pc1.eq}
\end{equation}
Energy generation equation
\begin{equation}
 \rhe[\vre\frac{d\epse}{d r}-\frac{\Pe}{\rhe}\{\frac{\vre}{\rhe}\frac{d\rhe}{d r} \}]=fQ_{\rm e}^+,
 \label{eg1.eq}
\end{equation}
where, $Q_{\rm e}^+=\trpe^2/\eta_{\rm e}$ and 
subscript `${\rm e}$' represents the quantities for accretion flow on the equatorial plane.
Here, $\trpe=\eta_{\rm e}\left({d \vpe}/{d r}-{\vpe}/{r}\right)=\eta_{\rm e}r(d\Omega/dr)$, 
$\eta_{\rm e}=\alpha_1\Pe/\omegk=\alpha_1\rhe a_{\rm se}^2/(\geff\omegk)$ and the
definition of  adiabatic sound speed from equation (\ref{eos.eq}) obtained as
$a_{\rm se}=\sqrt{\geff\Pe/\rhe}$. 
Integrating equations (\ref{ce1.eq}) and (\ref{pc1.eq}) become respectively, 
we assumed wedge accretion flow with `$\theta_{\rm e}$' angle around the equatorial plane,
\begin{equation}
 \dot{M}_{\rm in}=-4\pi r^2\rhe \vre {\rm cos\theta_e},
 \label{mar.eq}
\end{equation}
and assumed $\lambda\rightarrow\lambda_0$ as matter approaches to $r\rightarrow\rs$ and $\trp|_{horizon}=0$ with 
the help of equation (\ref{mar.eq}),
\begin{equation}
 \trpe=-\frac{\rhe \vre(\lambda_{\rm e}-\lambda_0)}{r},
 \label{trpc.eq}
\end{equation}
where, $\lambda_{\rm e}=r\vpe$ and $\lambda_0$ are specific angular momentum of the flow and 
specific angular momentum at the horizon, respectively. 
Here we assumed constant $\theta_{\rm e}$ with value of $\pi/3$ from the rotation axis,  
which is close to the disk thickness of almost all the results of this paper.
Integrating equation (\ref{rc1.eq}) with the help of equations (\ref{pc1.eq} and \ref{eg1.eq}), we get energy constant,
\begin{equation}
 {\cal E}=\frac{\vre^2}{2}+h_{\rm e}-\frac{\lambda_{\rm e}^2}{2r^2}+\frac{\lambda_{\rm e}\lambda_0}{r^2}-\int\Lambda_{\rm e}^-dr+\Phi,
 \label{engc.eq}
\end{equation}
This is known as the generalized specific energy of the flow \citep{kc14} and is a constant of motion for dissipative advective flows even in presence of cooling. 
Here, $\Lambda_{\rm e}^-=(1-f)(\lambda_{\rm e}-\lambda_0)(d\Omega/dr)$ and $h_{\rm e}=\epsilon_{\rm e}+\frac{P_{\rm e}}{\rho_{\rm e}}$ is specific enthalpy of the flow.
If we use $f=1$ then above equation becomes,
\begin{equation}
 E=\frac{\vre^2}{2}+h_{\rm e}-\frac{\lambda_{\rm e}^2}{2r^2}+\frac{\lambda_{\rm e}\lambda_0}{r^2}+\Phi,
 \label{engc1.eq}
\end{equation}
This is known as grand specific energy of the flow \citep{gl04,bdl08,kc13,kcm14} and is a constant of motion for the viscous flow.
If we again take inviscid flow then $\lambda_{\rm e}=\lambda_0$, so above equation becomes,
\begin{equation}
 B_{\rm e}=\frac{\vre^2}{2}+h_{\rm e}+\frac{\lambda_{\rm e}^2}{2r^2}+\Phi,
 \label{be.eq}
\end{equation}
This is the local specific energy of the flow and known as the Bernoulli parameter. 
We can use these energy constants for calculation of flow variables at critical point with two critical point conditions \citep{kscc13,kc17} 
or at horizon with a few assumptions
in order to find flow variables close to the horizon \citep{bdl08,kc13,kcm14,kc14,ck16}. 
Now, simplifying equations (\ref{rc1.eq}), (\ref{eg1.eq}) and (\ref{trpc.eq}) with the help of equations (\ref{ce1.eq}), (\ref{eos.eq}) 
and using expression of $\trpe$, we get,
\begin{equation}
\frac{d\Thre}{dr}=-\frac{\beta\tt}{N_{\rm eq}}\left[\frac{a_{\rm se}^2}{\geff}\left(\frac{1}{\vre}\frac{d\vre}{dr}+\frac{2}{r}\right)+f\Lambda_e^+\right],
 \label{dt1.eq}
\end{equation}
where, $\Lambda_{\rm e}^+={Q_{\rm e}^+}/{(\rhe \vre)}$.
\begin{equation}
 \frac{d\vre}{dr}=\frac{\frac{\vpe^2}{r}+F_r+2\frac{a_{\rm se}^2}{r}+
 \frac{f}{N_{\rm eff}}\Lambda_{\rm e}^+}{\vre-\frac{a_{\rm se}^2}{\vre}}
 =\frac{\cal N}{\cal D},
 \label{dv.eq}
\end{equation}
\begin{equation}
 \frac{d\lambda_{\rm e}}{dr}=\frac{2\lambda_{\rm e}}{r}+r^2\frac{d\Omega}{dr}~~ \mbox{and}~~ 
 \frac{d\Omega}{dr}=-\frac{\geff\vre\omegk (\lambda_{\rm e}-\lambda_0)}{\alpha_1a_{\rm se}^2r^2},
 \label{dl.eq}
\end{equation}
To find complete accretion solutions we have to solve all three differential equations (\ref{dt1.eq} - \ref{dl.eq}) with using specified flow parameters, namely,
${\cal E}, \lambda_0, \alpha_1, \beta$ and $\gamma$. Since BH accretion is transonic in nature, therefore, we have to find the location of sonic point but 
for dissipative flow sonic location is not known a priori. 
The equations and detail methodology for sonic point calculation have explained in section \ref{sec:solnproc} 
and appendix (\ref{sec:solodes}), respectively.

\section{Steps for solving ODEs}
\label{sec:solodes}
{
In order to get the complete inflow-outflow structure from our model equations, first we have to find the critical point (CP) for the 
transonic ADAF solution. Here we used the Runge-Kutta $4th$ order method to solve the differential equations along $r-$ and 
$\theta-$ direction.
The whole solution procedure is divided into the following steps,

{\it Step-1 Critical point location:} For given flow parameters $E$ (or ${\cal E}$), $\lambda_0$, $\alpha_1$, $\gamma$ and $\beta$, 
we obtained CP from iteration method by changing $\delta$ in following two parts. 

{\it Part-I Obtaining $\lmda, \vre, \Thre$ at $\rin$:} When we combined equations (\ref{frob.eq}) and (\ref{engc1.eq}) with 
the value of $\Lambda$ and expression of $\zeta$ at $\rin=1.001$. 
Thus we got a polynomial in $\ase$ or $\Thre$. 
Now, supplying the parameters ${E}$, $\lambda_0$, $\alpha_1$, $\gamma$ and $\beta$ then 
we solved the polynomial for $\Thre$ at $\rin$ for first choice $\delta=1$. 
Once $\Thre$ obtained at $\rin$, other quantities $\vre$ and $\lmda$ easily get with the help of $v_{\rm ff}$ and equation (\ref{frob.eq}).

{\it Part-II Finding $r_{\rm c}$:} We now can integrate differential equations (\ref{dt1.eq}-\ref{dl.eq}) outward 
from $\rin$ by using $\Thre$, $\vre$ and $\lmda$ and 
simultaneously, checking the sonic point equations (\ref{cn.eq}-\ref{cd.eq}). 
If sonic conditions are not satisfied then we reduced the value of $\delta<1$ and 
repeat the whole {\it part-I}. 
This solution procedure repeated till satisfying sonic conditions. 
When ensuring it then we obtained critical point location ($r_{\rm c}$) for given flow parameters. 

{\it Step-2 ADAF solution:} Once $r_{\rm c}$ obtained, we integrated equations (\ref{dt1.eq}-\ref{dl.eq}) outward along the 
radial direction for a given $\lambda_0$ with other disk parameters. 
Then we investigated outer boundaries of ADAF solution \citep{nkh97,lgy99} again by iteration method and changing $\lambda_0$ 
with repeating whole {\it step-1}.
Once $\lambda_0$ obtained for the ADAF solution and corresponding $r_{\rm c}$ then we went for the calculation of the 2D disk structure.

{\it Step-3 2D solution:} Here we supplied two more additional parameters $\alpha_2$ and $s$ for calculation of the polar flow variables 
of the disk structure with the outflows.  
We again divided procedure into two parts. 

{\it Part-A Obtaining $r_{\rm b}$:} When we obtained $r_{\rm c}$ for the ADAF solution. 
We integrated the radial fluid equations (\ref{dt1.eq}-\ref{dl.eq}) from $r_{\rm c}$ along $r-$ direction outward with 
some step size ($dr$) then at same $r$ again integrated the polar fluid equations (\ref{rc2.eq}-\ref{eg2.eq}, \ref{ce3.eq}) 
from $\theta=\pi/2$ along $\theta-$ direction towards rotation axis with some step size ($d\theta$). 
At each step size ($dr$) of $r$ we obtained the polar variables at $\theta=\pi/2$ from the equation (\ref{bcv.eq}) 
with the help of radial flow variables and it's derivatives of the ADAF. 
Now we again integrated radial equations at $r+dr$ then integrating polar equations from $\pi/2$ to $0$. 
These two integration run one by one till the radius $r=r_{\rm b}$, where the outflows or 2D disk solutions are existed, 
if not existed then stopped the integrations. 
Now we know the location of $r_{\rm b}$ then we can make a matrix 
for the density contour and velocity vectors plot, which is described in next part. 

{\it Part-B $N_{r}\times N_{\theta}$ Matrix:} We choose $N_{r}=N_{\theta}=256$ for the plotting of complete disk structure with using 
IDL (Interactive Data Language). The radial distance from $\rs$ to $r_{\rm b}$ and angular distance $\theta=\pi/2$ to $0$ are divided 
into $256$ parts and obtained radial and polar step size for integration, 
\eg radial integration step size $dr=(\rs-r_{\rm b})/(N_{r}-1)$ 
and polar integration step size $d\theta=\pi/2/(N_{\theta}-1)$. These two integration are run one by one as described in {\it part-A}, 
first along $r$ with one step size $dr$ then second along $\theta$ upto $\theta=0$ at same $r$. 
Next we increase $r$ by size $dr$ then we repeat same integrations and 
repeatedly doing so upto $r_{\rm b}$ then we get all matrix elements and therefore the complete 2D disk structure. 
}

\end{document}